\pdfoutput=1

\documentclass[12pt,a4paper]{article}

\usepackage{ifthen} 
\newboolean{pdflatex}
\setboolean{pdflatex}{true} 

\newboolean{articletitles}
\setboolean{articletitles}{true} 

\newboolean{uprightparticles}
\setboolean{uprightparticles}{false} 

\newboolean{inbibliography}
\setboolean{inbibliography}{false} 

\newboolean{wordcount}
\setboolean{wordcount}{false} 

\newboolean{prlversion}
\setboolean{prlversion}{false}

\def\paperauthors{LHCb collaboration} 
\def\paperasciititle{Measurement of antiproton production in pHe collisions at sqrt(sNN)=110 GeV} 
\def\papertitle{Measurement of antiproton production in \pHe collisions at $\sqsnn=110$\,GeV} 
\def\paperkeywords{{LHCb}, {LHC fixed target}, {cosmic ray antiprotons}} 
\def\papercopyright{\the\year\ CERN for the benefit of the LHCb collaboration} 
\def\paperlicence{CC-BY-4.0 licence}
\def\paperlicenceurl{https://creativecommons.org/licenses/by/4.0/}

\usepackage[top=1in, bottom=1.25in, left=1in, right=1in]{geometry}

\usepackage{microtype}
\usepackage{lineno}  
\usepackage{xspace} 
\usepackage{caption} 

\usepackage{graphicx}  
\usepackage{color}
\usepackage{colortbl}
\graphicspath{{./figs/}} 
\DeclareGraphicsExtensions{.pdf,.PDF,png,.PNG}

\usepackage{amsmath} 
\usepackage{amssymb}
\usepackage{amsfonts}
\usepackage{upgreek} 

\newcommand*\patchAmsMathEnvironmentForLineno[1]{%
\expandafter\let\csname old#1\expandafter\endcsname\csname #1\endcsname
\expandafter\let\csname oldend#1\expandafter\endcsname\csname
end#1\endcsname
 \renewenvironment{#1}%
   {\linenomath\csname old#1\endcsname}%
   {\csname oldend#1\endcsname\endlinenomath}%
}
\newcommand*\patchBothAmsMathEnvironmentsForLineno[1]{%
  \patchAmsMathEnvironmentForLineno{#1}%
  \patchAmsMathEnvironmentForLineno{#1*}%
}
\AtBeginDocument{%
\patchBothAmsMathEnvironmentsForLineno{equation}%
\patchBothAmsMathEnvironmentsForLineno{align}%
\patchBothAmsMathEnvironmentsForLineno{flalign}%
\patchBothAmsMathEnvironmentsForLineno{alignat}%
\patchBothAmsMathEnvironmentsForLineno{gather}%
\patchBothAmsMathEnvironmentsForLineno{multline}%
\patchBothAmsMathEnvironmentsForLineno{eqnarray}%
}

\usepackage{hyperxmp}

\usepackage[pdftex,
            pdfauthor={\paperauthors},
            pdftitle={\paperasciititle},
            pdfkeywords={\paperkeywords},
            pdfcopyright={Copyright (C) \papercopyright},
            pdflicenseurl={\paperlicenceurl}]{hyperref}

\usepackage[all]{hypcap} 

\setboolean{uprightparticles}{true} 
\usepackage{xspace} 
\usepackage{upgreek}

\def\lhcb {\mbox{LHCb}\xspace}

\def\MagUp {\mbox{\em Mag\kern -0.05em Up}\xspace}

\ifthenelse{\boolean{uprightparticles}}%
{

 \def\Ppi         {\ensuremath{\uppi}\xspace}

 \def\Pphi        {\ensuremath{\upphi}\xspace}

 \def\PDelta      {\ensuremath{\Delta}\xspace}                 
 \def\PXi      {\ensuremath{\Xi}\xspace}                 
 \def\PLambda      {\ensuremath{\Lambda}\xspace}                 
 \def\PSigma      {\ensuremath{\Sigma}\xspace}                 
 \def\POmega      {\ensuremath{\Omega}\xspace}                 
 \def\PUpsilon      {\ensuremath{\Upsilon}\xspace}                 
 
 \def\PB      {\ensuremath{\mathrm{B}}\xspace}                 
                  
 \def\PD      {\ensuremath{\mathrm{D}}\xspace}

 \def\PK      {\ensuremath{\mathrm{K}}\xspace}

 \def\Pe      {\ensuremath{\mathrm{e}}\xspace}

 \def\Pi      {\ensuremath{\mathrm{i}}\xspace}

 \def\Pn      {\ensuremath{\mathrm{n}}\xspace}                 
                  
 \def\Pp      {\ensuremath{\mathrm{p}}\xspace}

}
{

 \def\Ppi         {\ensuremath{\pi}\xspace}

 \def\Pphi        {\ensuremath{\phi}\xspace}

 \mathchardef\PDelta="7101
 \mathchardef\PXi="7104
 \mathchardef\PLambda="7103
 \mathchardef\PSigma="7106
 \mathchardef\POmega="710A
 \mathchardef\PUpsilon="7107
                  
 \def\PB      {\ensuremath{B}\xspace}                 
                  
 \def\PD      {\ensuremath{D}\xspace}

 \def\PK      {\ensuremath{K}\xspace}

 \def\Pe      {\ensuremath{e}\xspace}

 \def\Pi      {\ensuremath{i}\xspace}

 \def\Pn      {\ensuremath{n}\xspace}                 
                  
 \def\Pp      {\ensuremath{p}\xspace}

}

\makeatletter
\ifcase \@ptsize \relax
  \newcommand{\miniscule}{\@setfontsize\miniscule{4}{5}}
\or
  \newcommand{\miniscule}{\@setfontsize\miniscule{5}{6}}
\or
  \newcommand{\miniscule}{\@setfontsize\miniscule{5}{6}}
\fi
\makeatother

\DeclareRobustCommand{\optbar}[1]{\shortstack{{\miniscule (\rule[.5ex]{1.25em}{.18mm})}
  \\ [-.7ex] $#1$}}

\def\en         {{\ensuremath{\Pe^-}}\xspace}   

\def\pion   {{\ensuremath{\Ppi}}\xspace}

\def\pip    {{\ensuremath{\pion^+}}\xspace}
\def\pim    {{\ensuremath{\pion^-}}\xspace}
\def\pipm   {{\ensuremath{\pion^\pm}}\xspace}

\def\kaon    {{\ensuremath{\PK}}\xspace}
  \def\Kbar    {{\kern 0.2em\overline{\kern -0.2em \PK}{}}\xspace}

\def\KorKbar    {\kern 0.18em\optbar{\kern -0.18em K}{}\xspace}

\def\Kp      {{\ensuremath{\kaon^+}}\xspace}
\def\Km      {{\ensuremath{\kaon^-}}\xspace}

\def\Kmp     {{\ensuremath{\kaon^\mp}}\xspace}
\def\KS      {{\ensuremath{\kaon^0_{\mathrm{ \scriptscriptstyle S}}}}\xspace}

  \def\Dbar    {{\kern 0.2em\overline{\kern -0.2em \PD}{}}\xspace}
\def\D       {{\ensuremath{\PD}}\xspace}

\def\DorDbar    {\kern 0.18em\optbar{\kern -0.18em \PD}{}\xspace}

\def\Dstarpm {{\ensuremath{\D^{*\pm}}}\xspace}

\def\Bbar    {{\ensuremath{\kern 0.18em\overline{\kern -0.18em \PB}{}}}\xspace}

\def\BorBbar    {\kern 0.18em\optbar{\kern -0.18em B}{}\xspace}

  \def\Y#1S{\ensuremath{\PUpsilon{(#1S)}}\xspace}

\def\proton      {{\ensuremath{\Pp}}\xspace}
\def\antiproton  {{\ensuremath{\overline \proton}}\xspace}
\def\neutron     {{\ensuremath{\Pn}}\xspace}

\def\Lz          {{\ensuremath{\PLambda}}\xspace}
\def\Lbar        {{\ensuremath{\kern 0.1em\overline{\kern -0.1em\PLambda}}}\xspace}
\def\LorLbar    {\kern 0.18em\optbar{\kern -0.18em \PLambda}{}\xspace}

\def\to                 {\ensuremath{\rightarrow}\xspace}

\def\AT#1     {\ensuremath{A_{\mathrm{T}}^{#1}}\xspace}           

\def\C#1      {\ensuremath{\mathcal{C}_{#1}}\xspace}                       
\def\Cp#1     {\ensuremath{\mathcal{C}_{#1}^{'}}\xspace}                    
\def\Ceff#1   {\ensuremath{\mathcal{C}_{#1}^{\mathrm{(eff)}}}\xspace}        
\def\Cpeff#1  {\ensuremath{\mathcal{C}_{#1}^{'\mathrm{(eff)}}}\xspace}       
\def\Ope#1    {\ensuremath{\mathcal{O}_{#1}}\xspace}                       
\def\Opep#1   {\ensuremath{\mathcal{O}_{#1}^{'}}\xspace}                    

\newcommand{\tev}{\ifthenelse{\boolean{inbibliography}}{\ensuremath{~T\kern -0.05em eV}}{\ensuremath{\mathrm{\,Te\kern -0.1em V}}}\xspace}
\newcommand{\gev}{\ensuremath{\mathrm{\,Ge\kern -0.1em V}}\xspace}
\newcommand{\mev}{\ensuremath{\mathrm{\,Me\kern -0.1em V}}\xspace}
\newcommand{\kev}{\ensuremath{\mathrm{\,ke\kern -0.1em V}}\xspace}
\newcommand{\ev}{\ensuremath{\mathrm{\,e\kern -0.1em V}}\xspace}
\newcommand{\gevc}{\ensuremath{{\mathrm{\,Ge\kern -0.1em V\!/}c}}\xspace}
\newcommand{\mevc}{\ensuremath{{\mathrm{\,Me\kern -0.1em V\!/}c}}\xspace}
\newcommand{\gevcc}{\ensuremath{{\mathrm{\,Ge\kern -0.1em V\!/}c^2}}\xspace}
\newcommand{\gevgevcccc}{\ensuremath{{\mathrm{\,Ge\kern -0.1em V^2\!/}c^4}}\xspace}
\newcommand{\mevcc}{\ensuremath{{\mathrm{\,Me\kern -0.1em V\!/}c^2}}\xspace}

\def\mum  {\ensuremath{{\,\upmu\mathrm{m}}}\xspace}

\def\invnb {\ensuremath{\mbox{\,nb}^{-1}}\xspace}

\newcommand{\chisq}{\ensuremath{\chi^2}\xspace}

\newcommand{\chisqip}{\ensuremath{\chi^2_{\text{IP}}}\xspace}

\def\gsim{{~\raise.15em\hbox{$>$}\kern-.85em
          \lower.35em\hbox{$\sim$}~}\xspace}
\def\lsim{{~\raise.15em\hbox{$<$}\kern-.85em
          \lower.35em\hbox{$\sim$}~}\xspace}

\def\pt         {\ensuremath{p_{\mathrm{ T}}}\xspace}
\def\ptot       {\ensuremath{p}\xspace}

\def\geant      {\mbox{\textsc{Geant4}}\xspace}

\def\tell1  {TELL1\xspace}
\def\ukl1   {UKL1\xspace}

\usepackage{cite} 
\usepackage{mciteplus}

\def\sqsnn{\ensuremath{\protect\sqrt{s_{\scriptscriptstyle\rm NN}}}\xspace}
\def\pHe{\ensuremath{\rm p He}\xspace}
\def\pp{\proton\proton}
\def\pbar{\antiproton}

\def\pe{\ensuremath{\proton\en}\xspace}

\def\pvz{\ensuremath{z_{\mathrm{\scriptscriptstyle PV}}}\xspace}
\def\pvpe{\ensuremath{z_{\pe}}\xspace}

\def\DLLppi{\ensuremath{\mathrm{DLL_{p\Ppi}}}\xspace}
\def\DLLpK{\ensuremath{\mathrm{DLL_{pK}}}\xspace}

\def\mubarn{\ensuremath{\mathrm{ \,\upmu b}}\xspace}
\def\dsigmadpdpt{\ensuremath{\mathrm{d}^2\sigma/\mathrm{d}p\,\mathrm{d}p_{\mathrm{T}}}\xspace}
\def\dsigmaOndpdpt{\ensuremath{\dfrac{\mathrm{d}^2\sigma}{\mathrm{d}p\,\mathrm{d}p_{\mathrm{\scriptscriptstyle T}}}}\xspace}
\def\xF{\ensuremath{x_{\mathrm{F}}}\xspace}

\usepackage{longtable} 

\begin{document}

\renewcommand{\thefootnote}{\fnsymbol{footnote}}
\setcounter{footnote}{1}


\begin{titlepage}
\pagenumbering{roman}

\vspace*{-1.5cm}
\centerline{\large EUROPEAN ORGANIZATION FOR NUCLEAR RESEARCH (CERN)}
\vspace*{1.5cm}
\noindent
\begin{tabular*}{\linewidth}{lc@{\extracolsep{\fill}}r@{\extracolsep{0pt}}}
\ifthenelse{\boolean{pdflatex}}
{\vspace*{-1.5cm}\mbox{\!\!\!\includegraphics[width=.14\textwidth]{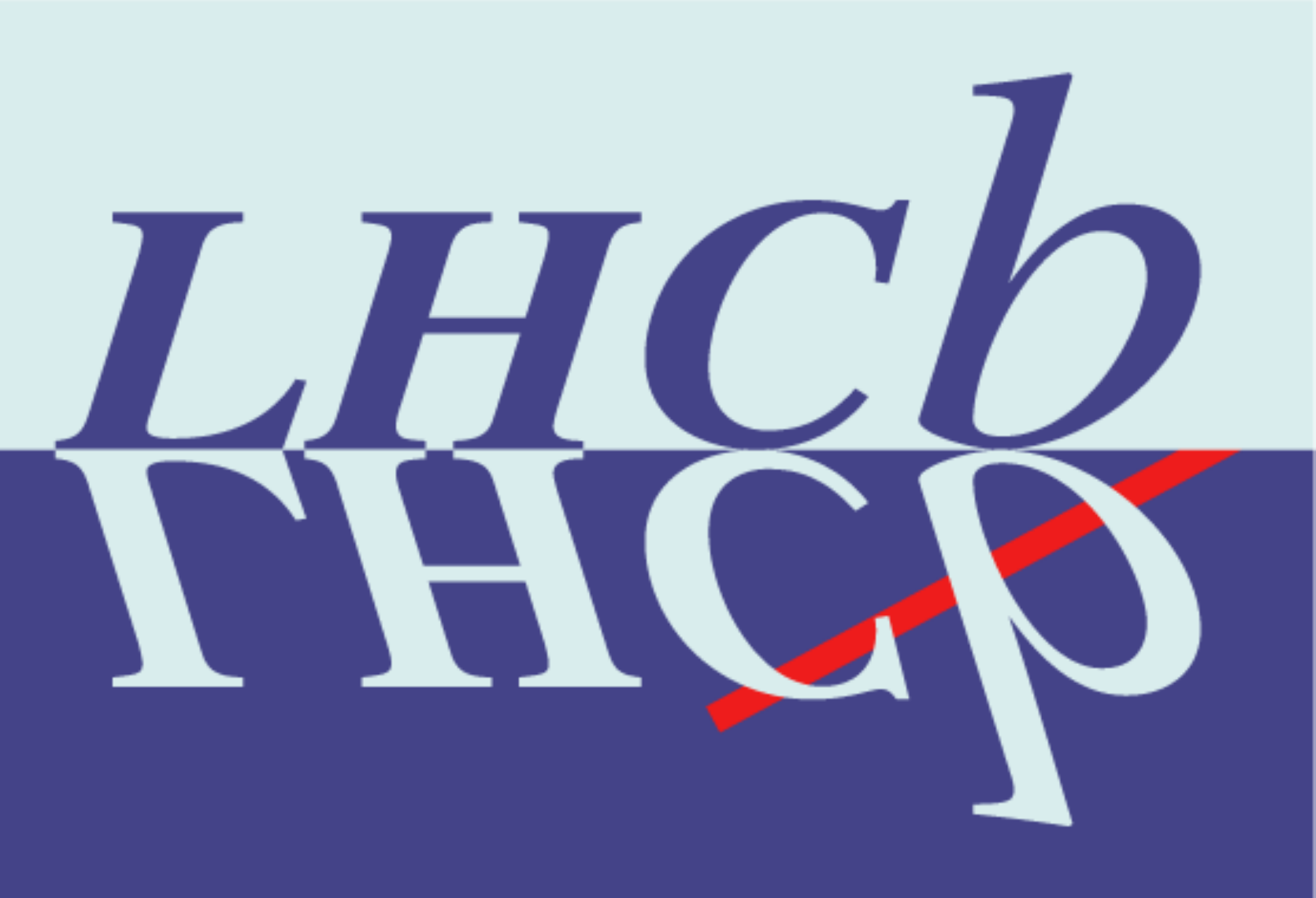}} & &}%
{\vspace*{-1.2cm}\mbox{\!\!\!\includegraphics[width=.12\textwidth]{lhcb-logo.eps}} & &}%
\\
 & & CERN-EP-2018-217 \\  
 & & LHCb-PAPER-2018-031 \\  
& & August 18, 2018 \\ 
 & & \\
\end{tabular*}

\vspace*{4.0cm}

{\normalfont\bfseries\boldmath\huge
\begin{center}

  \papertitle 
\end{center}
}

\vspace*{2.0cm}

\begin{center}
\paperauthors\footnote{Authors are listed at the end of this Letter.}
\end{center}

\vspace{\fill}

\begin{abstract}
  \noindent
 The cross-section for prompt antiproton production in collisions of
  protons with an energy of $6.5$\,TeV incident on helium nuclei at rest is measured 
  with the \lhcb experiment from a data set corresponding to an integrated luminosity of $0.5\invnb$.
  The target is provided by injecting helium gas into the LHC beam line
  at the \lhcb interaction point.
  The reported results, covering antiproton momenta between $12$ and $110\gevc$,
  represent the first direct determination of the antiproton production cross-section in \pHe collisions,
  and impact the interpretation of recent results on antiproton cosmic rays from space-borne experiments.

\end{abstract}

\vspace*{2.0cm}

\begin{center}
  Published in Phys.~Rev.~Lett. 121 (2018) 222001 
\end{center}

\vspace{\fill}

{\footnotesize 
\centerline{\copyright~\papercopyright. \href{\paperlicenceurl}{\paperlicence}.}}
\vspace*{2mm}

\end{titlepage}


\newpage
\setcounter{page}{2}
\mbox{~}

\cleardoublepage


\renewcommand{\thefootnote}{\arabic{footnote}}
\setcounter{footnote}{0}



\pagestyle{plain} 
\setcounter{page}{1}
\pagenumbering{arabic}


%

The antiproton fraction in cosmic rays
has been long recognized as a sensitive indirect probe for 
exotic astrophysical sources of antimatter, such as
dark matter annihilation~\cite{Gaisser:1974ks,Steigman:1976ev,Silk:1984zy,Stecker:1985jc,Hagelin:1985pv}. 
A substantial improvement in experimental accuracy for the measurement 
of the antiproton, \antiproton, over proton, \proton, flux ratio 
has recently been achieved by the
space-borne PAMELA~\cite{pamela} and AMS-02~\cite{ams} experiments.
Antiproton production in spallation of cosmic rays in the
interstellar medium, which is mainly composed of
hydrogen and helium, is expected to produce a \antiproton/\proton
flux ratio of $\mathcal{O}(10^{-4})$.
The observed excess of \pbar yields over current predictions 
for the known production sources~\cite{diMauro:2014zea,Giesen:2015ufa,Kappl:2015bqa,Reinert:2017aga}
can still be accommodated within the current uncertainties.
In the 10--100 GeV \pbar energy range, these uncertainties
are dominated by the limited knowledge of the \antiproton
production cross-section in the relevant processes. 
To date, no direct measurements of \pbar production in \pHe collisions have been made, 
and no data are available at a nucleon-nucleon center-of-mass (c.m.) energy of $\sqsnn\sim 100$\,GeV,  
relevant for the production of cosmic antiprotons above $10$\,GeV~\cite{Donato:2017ywo}.

This Letter reports the first measurement of prompt \antiproton
production in \pHe collisions carried out with the \lhcb experiment at
CERN using a proton beam with an energy of $6.5$\,TeV 
impinging on a helium gas target.
The forward geometry and particle identification (PID) capabilities of the \lhcb detector 
are exploited to reconstruct 
antiprotons with  momentum, \ptot, ranging from $12$ to $110\gevc$
and transverse momentum, \pt, between $0.4$ and $4.0\gevc$.
The integrated luminosity is determined
from the yield of elastically scattered atomic electrons.

The \lhcb detector is a single-arm forward
spectrometer covering the \mbox{pseudorapidity} range $2<\eta<5$, described in
detail in Refs.~\cite{Alves:2008zz,LHCb-DP-2014-002}, conceived for heavy-flavor
physics in \pp collisions at the CERN LHC.
The momentum of charged particles is measured to better than 1.0\% 
for $\ptot<110\gevc$.
The silicon-strip vertex locator (VELO), which surrounds the nominal \pp
interaction region, allows the measurement of the minimum distance of a track 
to a primary vertex (PV), the impact parameter (IP), 
with a resolution of $(15+29/\pt)\mum$, where \pt is in\,\gevc.
Different types of charged hadrons are distinguished using 
two ring-imaging Cherenkov detectors (RICH)~\cite{LHCb-DP-2012-003},
whose acceptance and performance define the \pbar kinematic range accessible
to this study. The first RICH detector has an inner acceptance limited to $\eta<4.4$ and
is used to identify antiprotons with momenta between $12$ 
and $60\gevc$.  The second detector
covers the range $3<\eta<5$ and can actively identify antiprotons with momenta between 
$30$ and $110\gevc$.
The scintillating-pad (SPD) detector and the electromagnetic calorimeter (ECAL)
included in the calorimeter system are also used in this study.

 The SMOG (System for Measuring Overlap with Gas) device~\cite{smog, LHCb-PAPER-2014-047} 
enables the injection of noble gases with pressure of $\mathcal{O}(10^{-7})$\,mbar
in the beam pipe section crossing  the VELO, 
allowing \lhcb to operate as a fixed-target experiment. 
This analysis is performed on data specifically acquired for this
measurement in May 2016. Helium gas was injected 
when the two beams circulating in the LHC accelerator~\cite{Evans:2008zzb} consisted 
of a small number, between $52$ and $56$, of proton bunches.
The proton-beam energy of $6.5$\,TeV corresponds
to $\sqsnn=110.5$\,GeV. In the proton-nucleon c.m. frame, the \lhcb acceptance
corresponds to central and backward rapidities $-2.8 < y^* < 0.2$,
and \pbar production can be studied for values of $x$-Feynman, the ratio of the 
\pbar longitudinal momentum to its maximal value, comprised between -0.24 and 0.

To avoid background from \pp collisions, the events used for this measurement were recorded 
when a bunch in the beam pointing toward \lhcb crosses the nominal 
interaction region without
a corresponding colliding bunch in the other beam. 
The online event selection consists of a hardware stage,
which requires activity in the SPD detector, and a software stage
requiring at least one reconstructed track in the VELO.
An unbiased control sample of randomly selected events 
is acquired independently of this online selection.

Simulated data samples are generated for \pHe collisions with
EPOS-LHC~\cite{Pierog:2013ria}, and for \pe normalization events with
ESEPP~\cite{Gramolin:2014pva}.  
The interaction of the generated particles with the detector, and its response,
are implemented using the \geant toolkit~\cite{Allison:2006ve,
*Agostinelli:2002hh} as described in Ref.~\cite{LHCb-PROC-2011-006}. 
Simulated collisions are uniformly distributed along the nominal beam
direction $z$ in the range $-1000 < z < +300$\,mm, where $z=0$\,mm 
is the nominal collision point.

Events with antiproton candidates must have a reconstructed
primary vertex within the fiducial region  $-700 < \pvz < +100$\,mm, where
high reconstruction efficiencies are achieved for both \pHe and \pe collisions.  
The PV position must be compatible with the beam profile and events
must have fewer than 5 tracks reconstructed in the VELO with negative pseudorapidity.
This selection is $(99.8 \pm 0.2)\%$  efficient 
for simulated reconstructed \pHe vertices, while suppressing vertices from
interactions with material, decays, and particle showers produced
in beam-gas collisions occurring upstream of the VELO.
The overlap of these backgrounds with a \pHe collision, an
effect not accounted for by the simulation,
causes an additional inefficiency of $(2.3 \pm 0.2)\%$, measured using the unbiased control sample.
The PV reconstruction efficiency for the signal events is estimated from
simulation and varies with \pvz from 66\% in the most
upstream region to 97\% around $\pvz=0$\,mm. 
This efficiency is sensitive to the 
PV track multiplicity, the angular distribution of primary tracks 
and the average position and profile of the beam.
Imperfections in these simulated distributions are accounted for
by weighting simulated events to improve the agreement with the distributions
observed in data. From the resulting variations of the PV reconstruction
efficiency, a relative systematic uncertainty is assigned, ranging from 
1.6\% to 3.3\%, depending on the \pbar kinematics.

Antiproton candidates are selected from
negatively charged tracks within the acceptance of at least one of the RICH detectors.
Additionally, \pbar candidates are required to
originate from the primary vertex by requiring \chisqip $<12$, where \chisqip is defined as 
the difference in the vertex-fit \chisq of the PV reconstructed with and
without the track under consideration.
The reconstruction efficiency for prompt antiprotons, $\epsilon_{\rm rec}$, including
the detector acceptance and the tracking efficiency, is determined
from simulation in three-dimensional bins of \ptot, \pt and \pvz. 
The width of the momentum bins increases as a power law of \ptot to have approximately an 
equal number of candidates in each of 18 bins.
Ten \pt bins are chosen with the same criterion, while 12 uniform bins are used in \pvz.
Bins in which $\epsilon_{\rm rec}$ is below 25\% are not used in order to reduce 
systematic uncertainties, effectively shortening the \pvz fiducial region for kinematic
bins at the edges of the detector acceptance.
The average value of $\epsilon_{\rm rec}$ in the remaining bins is 61\%.
The tracking efficiency obtained from
the simulation is corrected by a factor determined from calibration
samples in \pp-collision data. This correction factor is consistent with unity 
in all kinematic bins within its systematic uncertainty of $0.8\%$~\cite{LHCB-DP-2013-002}. 
The \pvz dependence of the tracking efficiency 
is checked using $\KS\to\pip\pim$  decays in the \pHe sample where one of the tracks is 
reconstructed without using VELO information.  No significant 
differences between data and simulation are observed. 
A systematic uncertainty, varying between 1.0\% and 4.0\% depending on
$\eta$, accounts for \pbar hadronic interactions in the detector
material, whose rate is known with 10\% accuracy~\cite{LHCB-DP-2013-002}.
The efficiency of the \chisqip requirement is parameterized as a function of \pt and \ptot, 
averaging to $96.1\%$, with a $1.0\%$ uncertainty from the parameterization accuracy.
The online selection efficiency is unity, within $10^{-5}$, as determined from the unbiased control sample.

\ifthenelse{\boolean{prlversion}}{
\begin{figure}[bt]
  \centering
  \includegraphics[width=0.99\linewidth]{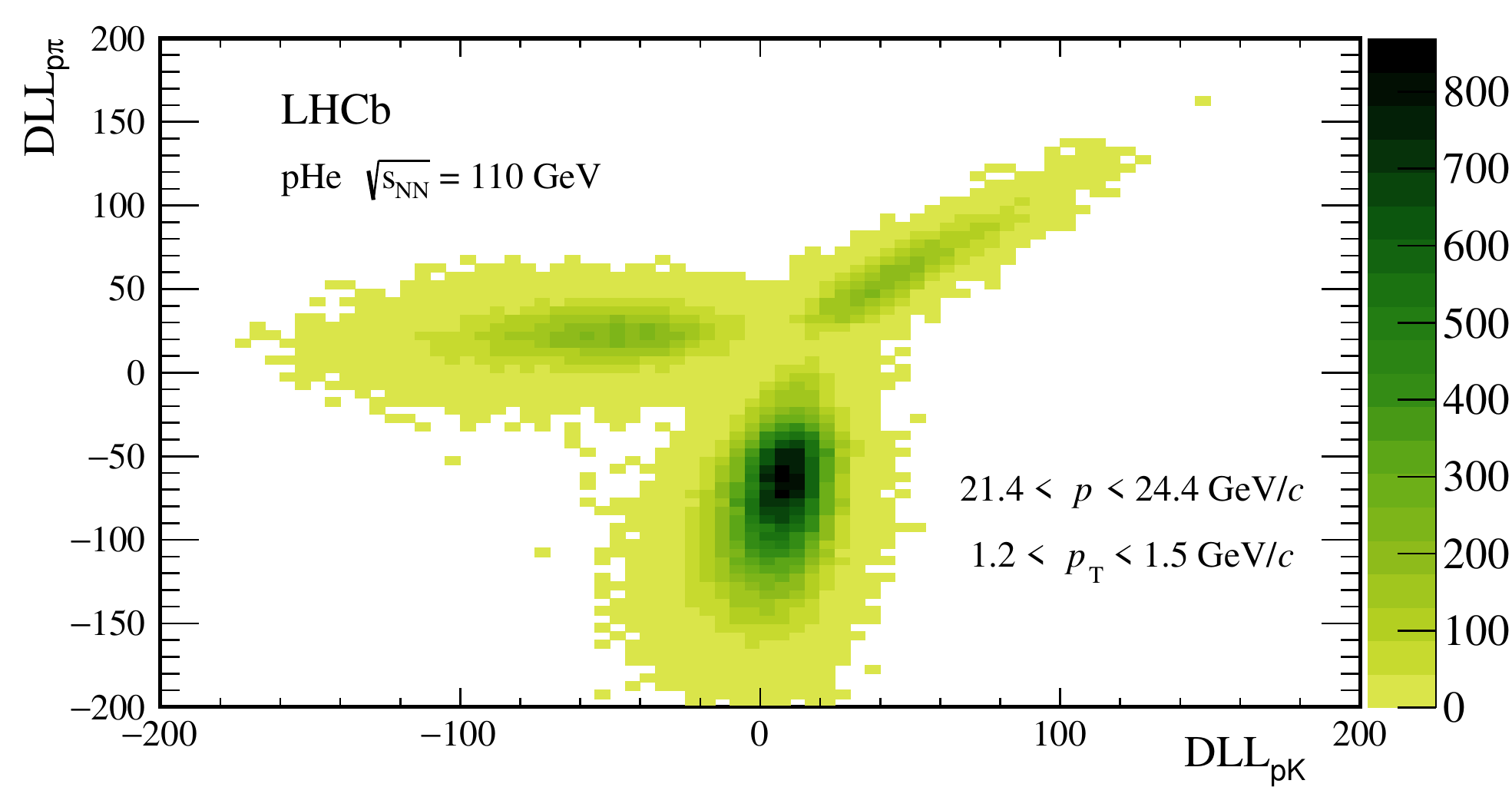}
  
  \includegraphics[width=0.99\linewidth]{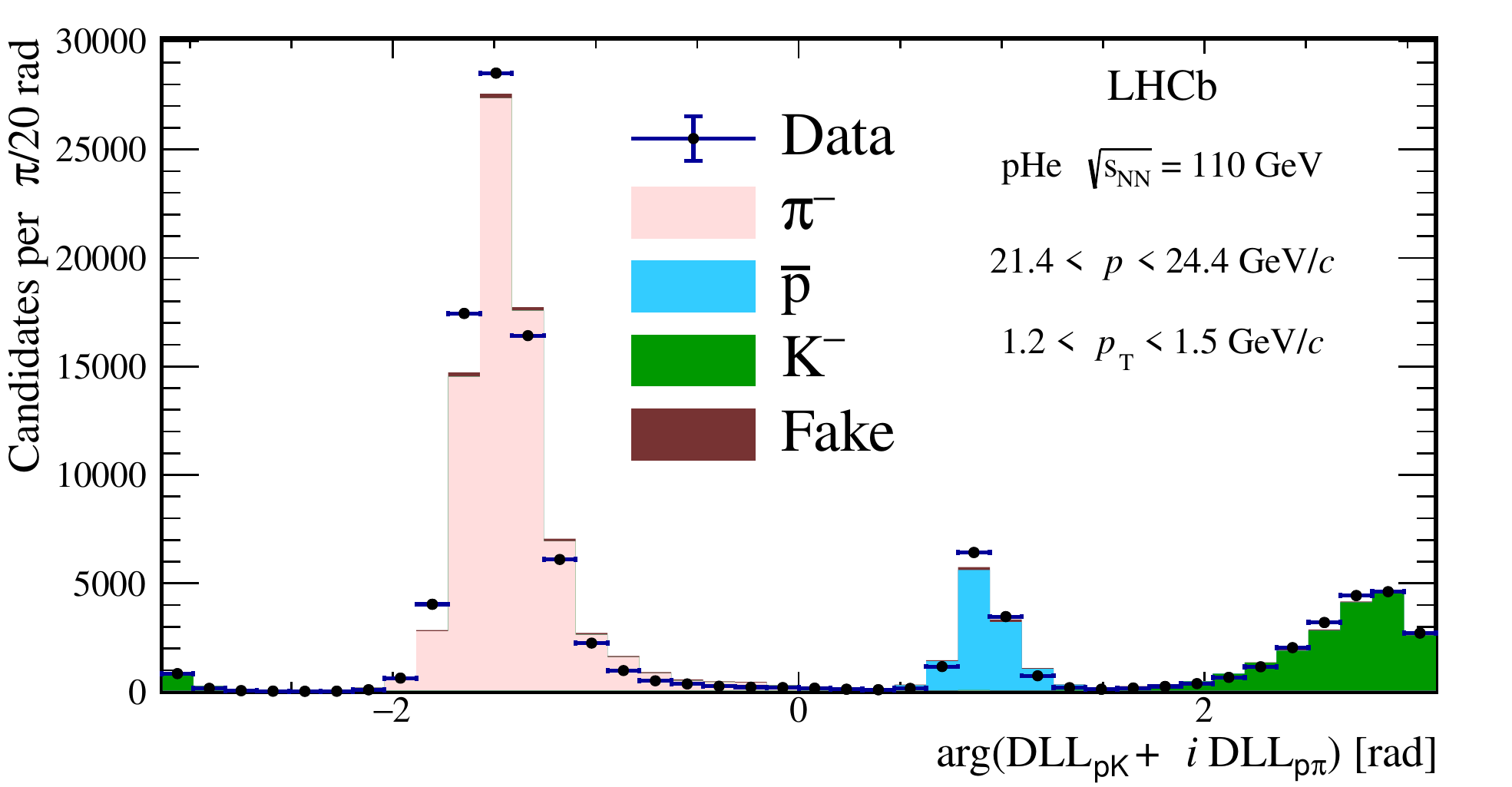}
  \caption{\label{fig:pidfitExample}
    Two-dimensional template fit to the PID distribution of negatively charged tracks for a particular bin (\mbox{$ 21.4 <
    \ptot < 24.4\gevc$}, \mbox{$1.2 < \pt < 1.5\gevc$}). 
    The (\DLLpK, \DLLppi) distribution, shown in the top plot, is fitted to determine the relative
    contribution of \pim, \Km and \pbar particles, using simulation to determine the template distributions and
    the fraction of fake tracks (which are barely visible).
    In the bottom plot, the result of the fit is projected into the variable \mbox{$\arg{(\DLLpK + i\, \DLLppi)}$}.}
\end{figure}}{}

Based on studies of simulated \pHe collisions,
the sample of negatively charged tracks is  dominated by 
\pim, \Km and \pbar hadrons. 
In a small fraction of cases, 1.7\% in the simulation,
tracks do not correspond to the trajectories of 
real charged particles and are labelled as fake tracks.
Particle identification is based on the response of the RICH
detectors, from which two quantities are determined: the difference between 
the log likelihood of the proton and pion hypotheses, \DLLppi, and
that between the proton and kaon hypotheses, \DLLpK~\cite{LHCb-DP-2012-003}. 
Three sets of templates for each particle species are determined from simulation,
from \pHe data, and from \pp data collected in 2016.
The \pHe calibration samples consist of selected $\KS\to\pip\pim$
decays for pions, $\Lz\to\proton\pim (\Lbar\to\antiproton\pip)$ for (anti)protons and
$\Pphi\to\Kp\Km$ for kaons. 
Calibration samples in \pp data also include $\Dstarpm\to\DorDbar^0(\Kmp\pipm)\pipm$ decays.
Simulation is used for the template of fake tracks.

Two methods are used to determine the  \pbar fraction in each kinematic bin:
a two-dimensional binned extended-maximum-likelihood fit, illustrated 
in Fig.~\ref{fig:pidfitExample}, and a cut-and-count method~\cite{LHCb-PAPER-2011-037},
which uses exclusive high-purity samples selected with tight requirements for each particle species.
\ifthenelse{\boolean{prlversion}}{}{
\begin{figure}[bt]
  \centering
  \includegraphics[width=0.99\linewidth]{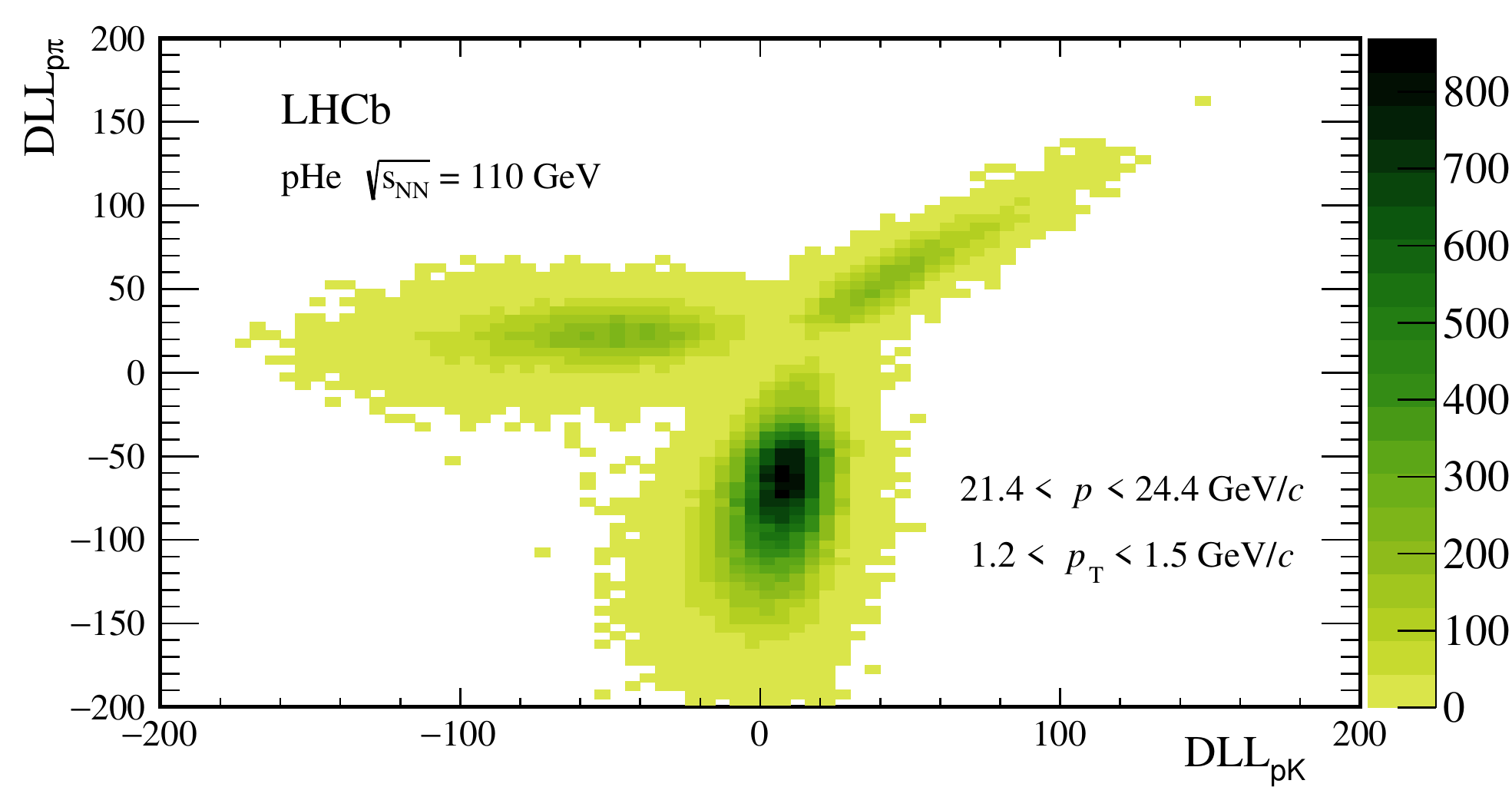}
  
  \includegraphics[width=0.99\linewidth]{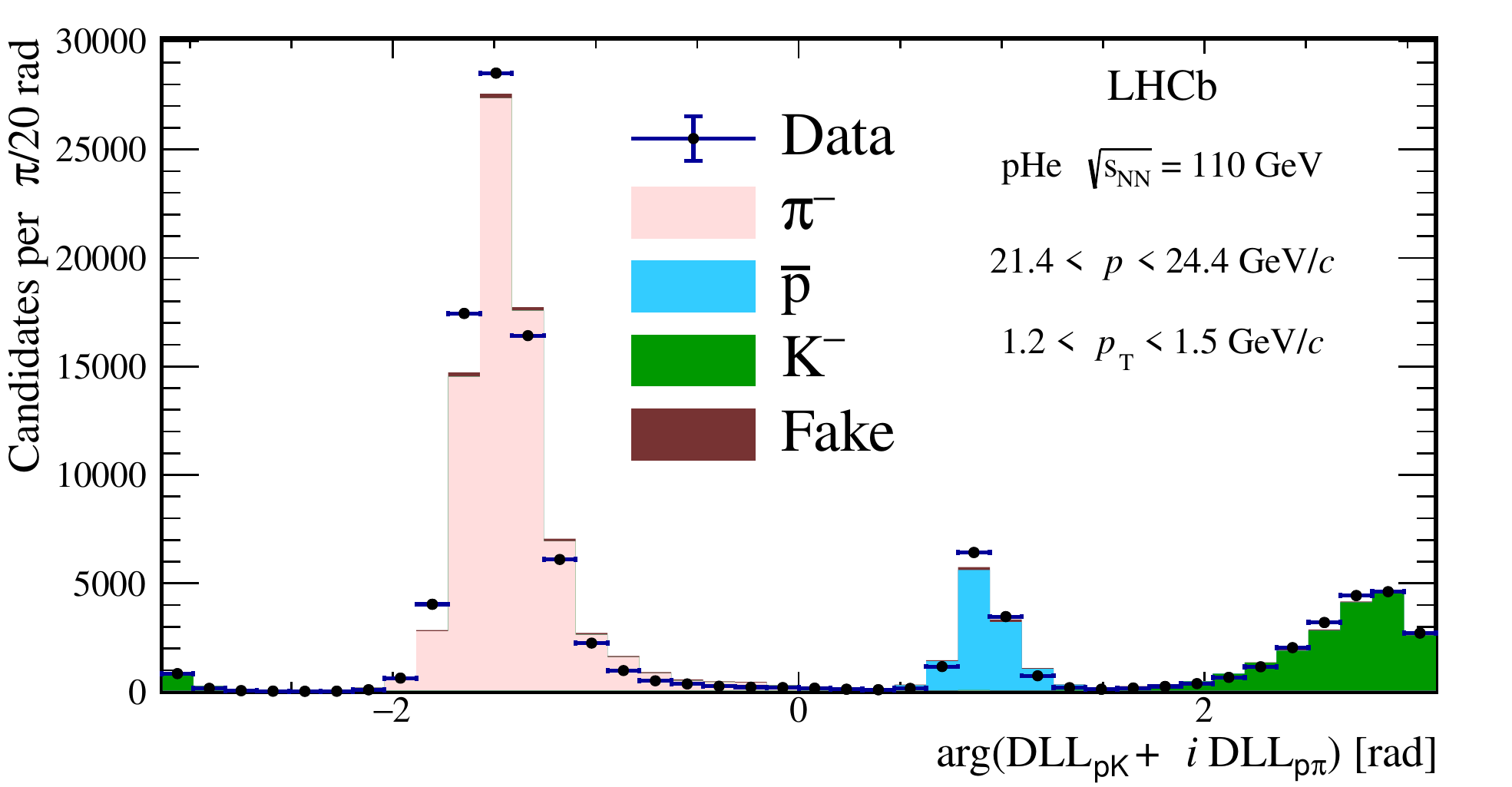}
  \caption{\label{fig:pidfitExample}
    Two-dimensional template fit to the PID distribution of negatively charged tracks for a particular bin (\mbox{$ 21.4 <
    \ptot < 24.4\gevc$}, \mbox{$1.2 < \pt < 1.5\gevc$}). 
    The (\DLLpK, \DLLppi) distribution, shown in the top plot, is fitted to determine the relative
    contribution of \pim, \Km and \pbar particles, using simulation to determine the template distributions and
    the fraction of fake tracks (which are barely visible).
    In the bottom plot, the result of the fit is projected into the variable \mbox{$\arg{(\DLLpK + i\, \DLLppi)}$}.}
\end{figure}}
The probability $P_{ij}$
that a candidate of species~$i$ is classified as species~$j$ is
obtained from the templates. The  $4\times 4$ $P_{ij}$ matrix is then inverted
to derive the yield of each particle species.
For each kinematic bin, the central value for the \pbar fraction is obtained from the 
average of the two methods using the templates from simulation, while half the difference
is used to estimate the systematic uncertainty. Bias from
the imperfections of the simulated RICH response, which are visible in
Fig.~\ref{fig:pidfitExample}, is estimated from the average differences 
among the results using the three available template sets, which are used 
to assign an additional uncertainty, correlated among bins.
The total uncertainty is typically a few percent, although
larger uncertainties affect the bins at the edges of the detector acceptance.

In the simulation, the non-prompt antiprotons surviving the \chisqip
requirement constitute a fraction of the selected \pbar sample 
varying between 1\% and 3\% depending on \pt. These are due to hyperon decays, in
90\% of cases, or secondary interactions.
This fraction is corrected by a factor
$1.5 \pm 0.3$, to account for differences between simulation and data as
determined in the region of the \chisqip distribution dominated by
hyperon decays. The resulting correction to the \pbar yield averages to  $-2.4\%$.

Collisions on the residual gas in the LHC beam vacuum, with a pressure
of $\mathcal{O}(10^{-9})$\,mbar and unknown composition,   
can contribute to the \pbar yield.
Residual-gas analysis, performed in the absence of beam,
indicates that the contamination is $\mathcal{O}(1)\%$ and is dominated by hydrogen. 
To evaluate this background source, including a possible beam-induced component,
a control sample of beam-gas collisions was acquired before injection of the helium gas.
Data collected with and without helium gas have the same vacuum pumping configuration
and thus identical residual gas composition and pressure.
The yield of selected events in data without helium gas, scaled according to the corresponding number of
protons on target, is subtracted from the result leading to an average correction of
$(-0.6 \pm 0.1)\%$, where the uncertainty accounts
for the background variation over time.
The average PV track multiplicity is found to be smaller
in collisions without injected gas, confirming that the residual gas is dominated by hydrogen.

Since the injected gas pressure is not precisely known,
the integrated luminosity of the data sample is determined from the yield of 
electrons from elastic scattering of the proton beam.
Scattered electrons are simulated in the polar angle range
$3<\theta<27$\,mrad, outside of which they cannot be reconstructed in
LHCb. The corresponding cross-section is calculated to be 
$184.8\pm 1.8\mubarn$~\cite{Gramolin:2014pva}, where the
uncertainty is due to the proton form factors and radiative corrections.
Scattered electrons are selected  from events with a 
single reconstructed track. The electron candidate is required to have
 $\ptot<15\gevc$, $\pt<0.12\gevc$, a polar angle in the
 range $11<\theta<21$\,mrad, and to originate from the fiducial region.
The longitudinal position of the
scattering vertex \pvpe is determined from the position of
minimum approach to the beam line, with a resolution of $9$\,cm.
The track reconstruction efficiency in the selected \pvpe and $\theta$ ranges
is determined from simulation to be $16.3\%$.
A loose requirement is placed on the energy deposited in the ECAL to 
identify the track as an electron.
Background events that could mimic this signature are expected to be    
mostly soft nuclear collisions where the initial nucleons do not dissociate, and the
detected particle is produced by a colorless exchange of gluons or photons.   
Since the products of this process must be charge-symmetric, the background
yield is determined from events with a single positron candidate.
\begin{figure}[tb]
  \centering
  \includegraphics[width=\linewidth]{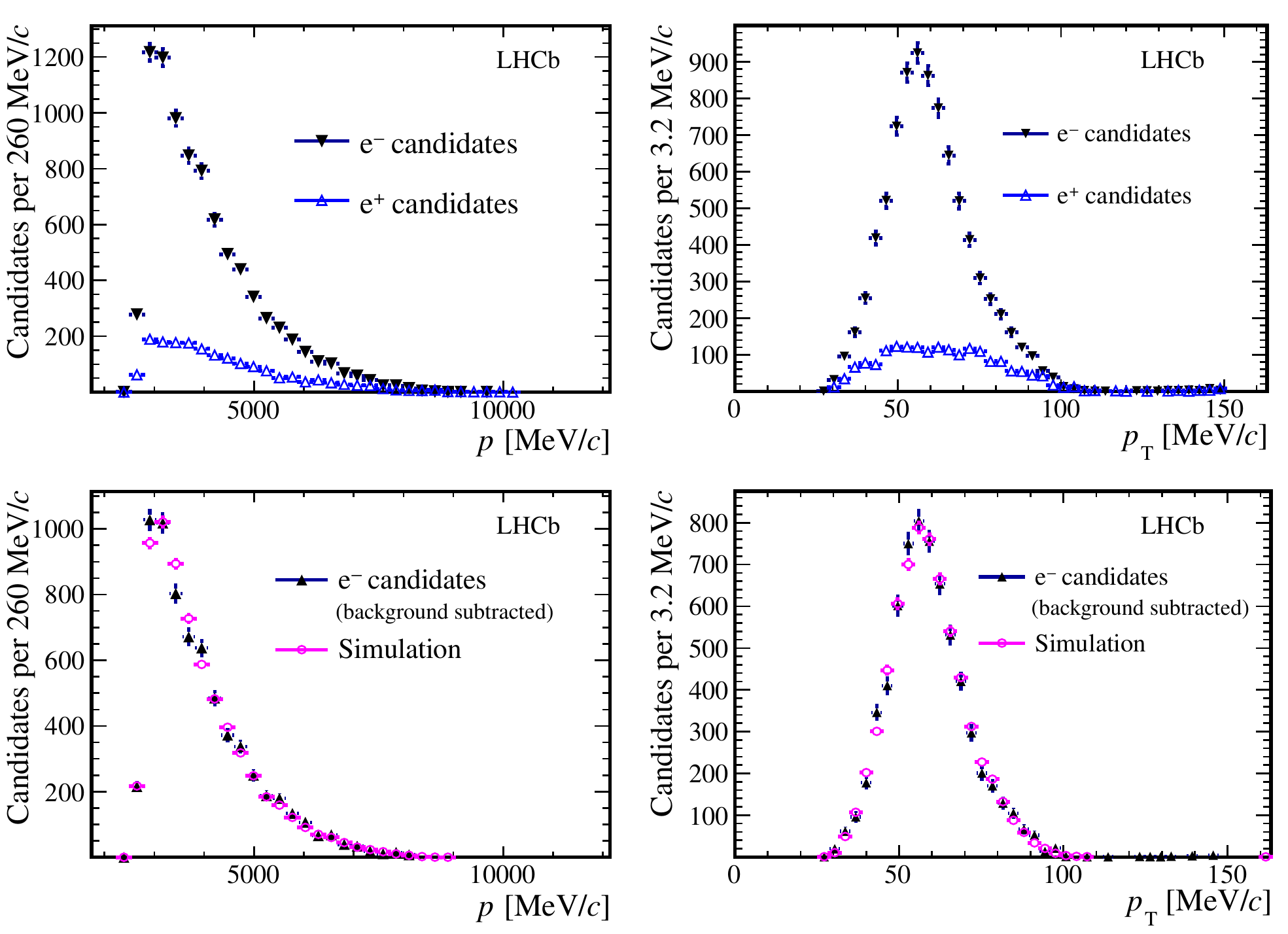}
  \caption{Distributions of (left) momentum and (right) transverse momentum for (top) single electron and
    single positron candidates, and (bottom) background-subtracted electron candidates, compared with the
    distributions in simulation, which are normalized to the data yield. }
  \label{fig:en_kin}
\end{figure}

Background is further suppressed by two multivariate classifiers, implemented 
using a BDT algorithm~\cite{AdaBoost}. The first exploits the geometric and 
kinematic properties of the
candidate electron. The second uses multiplicity variables
to veto any extra activity in the event. In both cases 
the classifiers are trained using \pe simulated events for the signal
and single-positron events from data for the background.  
Loose requirements are placed on the response of the BDT  discriminants, with a combined
efficiency of $96\%$ for simulated
\pe events. The overlap of a \pe event with another beam-gas
interaction causes an additional
inefficiency, measured to be $(9.4 \pm 0.7)\%$ in the unbiased control sample.
A possible charge asymmetry of the background, 
estimated from the EPOS simulation, leads to a systematic uncertainty of $1.9\%$. 
As is done for the \pbar candidates, the unbiased control events are used
to measure the online selection efficiency, $(98.3\pm 0.3)\%$,
and the data without helium gas are used to determine the contribution
from scattering on residual gas, $(1.0 \pm 0.3)\%$.

The momentum distributions of the selected candidates are shown in Fig.~\ref{fig:en_kin},
where a good agreement with the simulated \pe signal is observed after
background subtraction.
The low reconstruction efficiency, due to the fact that the observed electrons are predominantly
produced at the edges of the \lhcb acceptance and are subject to relevant 
 energy losses by bremsstrahlung when crossing the detector material,
is the major source of systematic uncertainty on the luminosity. 
The stability of the result is checked against additional
requirements on the  most critical variables, notably the number of
reconstructed VELO hits and the azimuthal angle, whose distribution is strongly affected
by the spectrometer magnetic field.
The largest variation of the result, a relative $5.0\%$, is assigned
as systematic uncertainty on the electron reconstruction efficiency. 
Taking also into account an uncertainty of $2.3\%$ from the beam 
and VELO simulated geometry, the total systematic
uncertainty on the luminosity is $6.0\%$.

The integrated \pHe luminosity is determined from the efficiency-corrected 
yield, divided by the product of the \pe cross-section and the helium atomic number.
Gas ionization effects are found to be negligible.
Avoiding any assumption on the $z$ dependence of the 
gas density, the integrated luminosity is calculated with 12 $\pvpe$-bins
across the fiducial region, resulting in 
$484 \pm 7 \pm 29\mubarn^{-1}$, where the first uncertainty is statistical and the
second is systematic.
From the knowledge of the number of delivered protons, the target gas pressure 
is found to be $2.6\times 10^{-7}$\,mbar, which is compatible with the expected helium pressure.

Table~\ref{tab:syst} presents the list of uncertainties on the \pbar cross-section measurement,
categorized into correlated and uncorrelated sources among kinematic bins.
\begin{table}[b]
  \centering
\caption{Relative uncertainties on the \pbar production cross-section. 
The ranges refer to the variation among kinematic bins. }
\label{tab:syst}  
\ifthenelse{\boolean{wordcount}}{}{
\begin{tabular}{llc}
\hline
\multicolumn{3}{l}{ Statistical}  \\
~~& \pbar yields  ~~~~     &  $0.5 - 11\%$ ($<2\%$ for most
bins) \\
& Luminosity     ~~~~~   &$1.5 - 2.3 \%$ \\
\multicolumn{2}{l}{ Correlated systematic}  \\
& Luminosity      ~~~~~  &   $6.0 \%$ \\
& Event and PV selection &$0.3 \%$ \\
& PV reconstruction      &   $0.4 - 2.9 \%$ \\
& Tracking               &   $1.3 - 4.1 \%$ \\
& Non-prompt background   &   $0.3 - 0.5 \%$  \\
& Target purity          &   $0.1 \%$ \\
& PID                    &   $3.0 - 6.0 \%$  \\
\multicolumn{2}{l}{ Uncorrelated systematic}   \\
& Tracking               &   $1.0 \%$ \\
& IP cut efficiency      &   $1.0 \%$ \\
& PV reconstruction      &   $1.6 \%$ \\
& PID                    &   $0 - 36 \%$  ($<5\%$ for most bins)  \\
& Simulated sample size  &   $0.4 - 11 \%$ ($<2\%$ for most bins)  \\
\hline
\end{tabular}
}
\end{table}
The correlated systematic uncertainty is dominated by the uncertainty on the
luminosity determination.  
The net effect of migration between kinematic bins due to resolution effects 
is found to be negligible.
A major difference between the fixed-target configuration and the
standard \pp-collision data taking in \lhcb is the extension of the luminous
region. As a consequence, the result is checked to be independent of \pvz 
within the quoted uncertainty in  all kinematic bins.
Furthermore, the results do not show any significant dependence
on the time of data taking.

\begin{figure}[ptb]
    \centering
    \includegraphics[width=\linewidth]{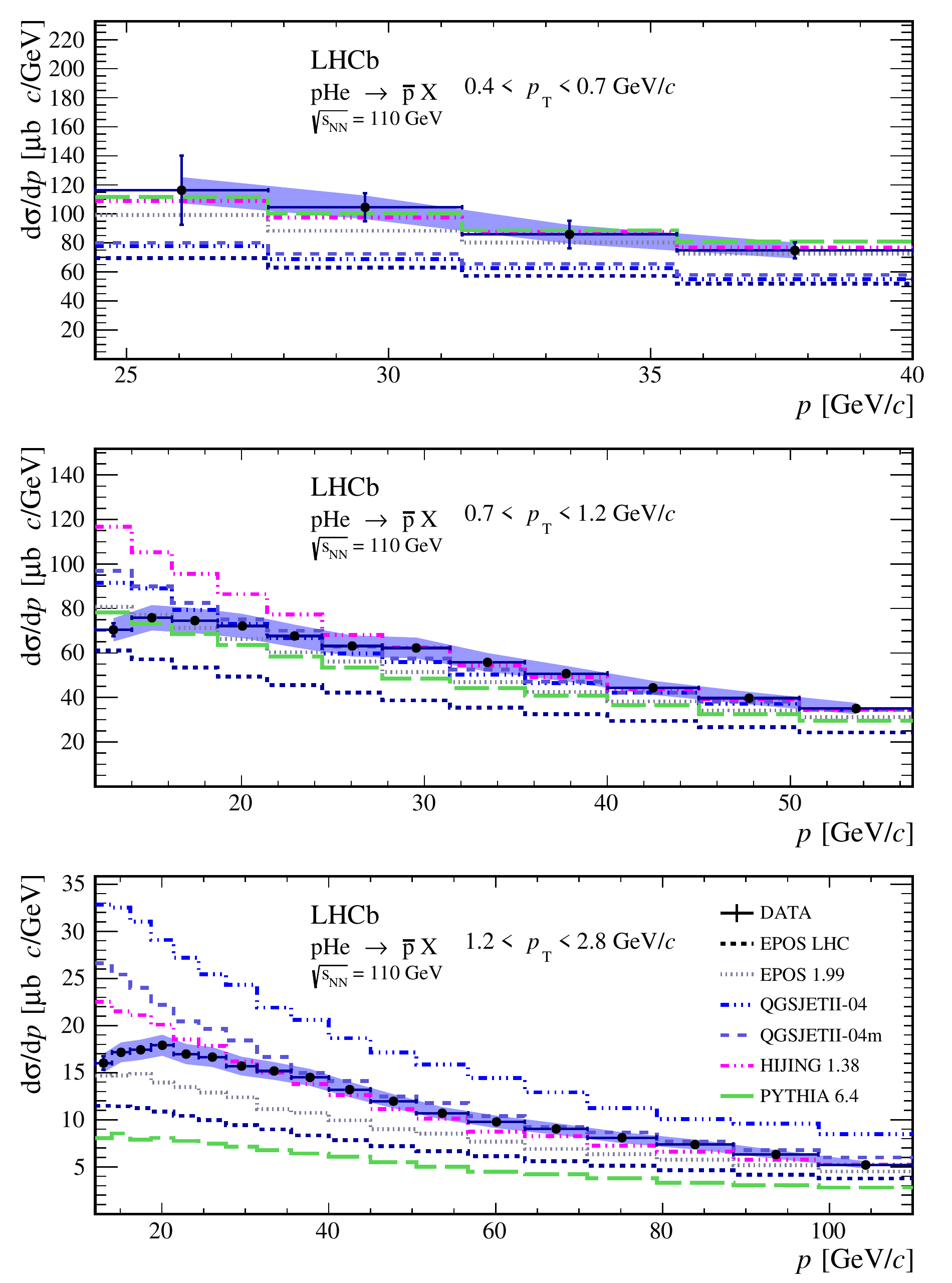}
    \caption{Antiproton production cross-section per He nucleus as a function of momentum,
      integrated over various \pt regions. The data points
      are compared with predictions from theoretical models.
      The uncertainties on the data points  are uncorrelated only, while the shaded area 
      indicates the correlated uncertainty.
    }
    \label{fig:result}
\end{figure}

The \pbar production cross-section 
is determined in each kinematic bin from a sample of $33.7$ million
reconstructed \pHe collisions, yielding $1.5$ million antiprotons as
determined from the PID analysis.
In Fig.~\ref{fig:result}, the results, integrated in different kinematic regions,
are compared with the prediction of several models: EPOS-LHC~\cite{Pierog:2013ria}, the pre-LHC EPOS 
version 1.99~\cite{Pierog:2009zt}, HIJING 1.38~\cite{Gyulassy:1994ew}, the QGSJET model 
II-04~\cite{Ostapchenko:2010vb} and its low-energy extension QGSJETII-04m, motivated 
by \pbar production in cosmic rays~\cite{Kachelriess:2015wpa}. 
The results are also compared 
with the PYTHIA6.4~\cite{Sjostrand:2006za} prediction for 
\mbox{$2\times\left[\sigma(\pp\to\pbar X)+\sigma(\proton\neutron\to\pbar X)\right]$},
not including nuclear effects.
The shapes are well reproduced except at
low rapidity, and the absolute \pbar yields deviate by up to a factor of two.
Numerical values for the double-differential cross-section 
\dsigmadpdpt in each kinematic bin are available in \ifthenelse{\boolean{prlversion}}{
the Supplemental Material.}{Appendix~\ref{sec:Supplemental-App}.}

The total yield of \pHe inelastic collisions which are visible in \lhcb is 
determined from the yield of reconstructed primary vertices and is
found to be compatible with
EPOS-LHC:  $\sigma_{\rm vis}^{\rm LHCb}/ \sigma_{\rm vis}^{\rm EPOS-LHC}  = 
1.08 \pm 0.07 \pm 0.03$, where the first uncertainty is due to the luminosity and the second to 
the PV reconstruction efficiency. The result indicates that the significant excess of \pbar 
production over the EPOS-LHC prediction, visible in Fig.~\ref{fig:result},
is mostly due to the \pbar multiplicity.

In summary, using a \pHe collision data sample, corresponding to an integrated
luminosity of $0.5\invnb$,  the \lhcb collaboration has performed the first measurement of 
antiproton production in \pHe collisions.
The precision is limited by systematic effects and is better than a relative 
10\% for most kinematic bins, well below 
the spread among models describing
\pbar production in nuclear collisions.
The energy scale, $\sqsnn=110$\,GeV, and the
measured range of the antiproton kinematic spectrum 
are crucial for interpreting the precise \pbar cosmic ray measurements from the PAMELA and AMS-02 experiments
by improving the precision of the secondary \pbar cosmic ray flux prediction~\cite{Reinert:2017aga,Korsmeier:2018gcy}.


\section*{Acknowledgements}

\noindent  We are grateful to our colleagues from the cosmic ray community,
O.~Adriani, F.~Donato, L.~Bonechi and A.~Tricomi, for suggesting 
this measurement, to T.~Pierog and S.~Ostapchenko for their advice 
on the theoretical models for antiproton production, and
to B.~Ward and A.~V.~Gramolin for their advice on the model and uncertainty
for \pe scattering.
We express our gratitude to our colleagues in the CERN
accelerator departments for the excellent performance of the LHC. We
thank the technical and administrative staff at the LHCb
institutes.
We acknowledge support from CERN and from the national agencies:
CAPES, CNPq, FAPERJ and FINEP (Brazil); 
MOST and NSFC (China); 
CNRS/IN2P3 (France); 
BMBF, DFG and MPG (Germany); 
INFN (Italy); 
NWO (Netherlands); 
MNiSW and NCN (Poland); 
MEN/IFA (Romania); 
MSHE (Russia); 
MinECo (Spain); 
SNSF and SER (Switzerland); 
NASU (Ukraine); 
STFC (United Kingdom); 
NSF (USA).
We acknowledge the computing resources that are provided by CERN, IN2P3
(France), KIT and DESY (Germany), INFN (Italy), SURF (Netherlands),
PIC (Spain), GridPP (United Kingdom), RRCKI and Yandex
LLC (Russia), CSCS (Switzerland), IFIN-HH (Romania), CBPF (Brazil),
PL-GRID (Poland) and OSC (USA).
We are indebted to the communities behind the multiple open-source
software packages on which we depend.
Individual groups or members have received support from
AvH Foundation (Germany);
EPLANET, Marie Sk\l{}odowska-Curie Actions and ERC (European Union);
ANR, Labex P2IO and OCEVU, and R\'{e}gion Auvergne-Rh\^{o}ne-Alpes (France);
Key Research Program of Frontier Sciences of CAS, CAS PIFI, and the Thousand Talents Program (China);
RFBR, RSF and Yandex LLC (Russia);
GVA, XuntaGal and GENCAT (Spain);
the Royal Society
and the Leverhulme Trust (United Kingdom);
Laboratory Directed Research and Development program of LANL (USA).

\clearpage
\ifthenelse{\boolean{prlversion}}{\section{Supplemental material for LHCb-PAPER-2018-031}}{
\appendix
\section{Numerical results}
}
\label{sec:Supplemental-App}

The numerical results for the antiproton production cross-section per He nucleus
in \pHe collisions at $\sqsnn=110$\,GeV are reported in
Table~\ref{tab:results} for each kinematic bin.

The cross-section for \pHe inelastic collisions whose primary vertex can be reconstructed
in \lhcb (at least three primary tracks within the acceptance of the VELO detector)
is measured to be
$$ \sigma_{\rm vis}^{\rm LHCb} = ( 71.9 \pm 4.5 \pm 2.3)\,\text{mb},  $$ 
where the first uncertainty is due to the luminosity and the second to 
the reconstruction efficiency. The EPOS-LHC prediction is $66.6$\,mb for this visible cross-section,
and $118$\,mb for the total inelastic cross-section.
The fraction of events not reconstructible in \lhcb varies between $33$ and $44$\%
among the EPOS-LHC, QGSJETII-04 and HIJING models.

\begin{table}[h]
\caption{Numerical results for the
measured prompt \pbar production cross-section. 
The reported values are the double-differential cross-section \dsigmadpdpt  
per He nucleus in the laboratory frame, averaged over the given kinematic range of each bin. 
The uncertainty is split into an uncorrelated uncertainty $\delta_{\text{uncorr}}$, and 
an uncertainty $\delta_{\text{corr}}$ which is fully correlated among the kinematic bins.
For both uncertainties, the systematic uncertainty, dominant for most bins, and 
the statistical uncertainty, are added in quadrature. 
The average value within each bin is also reported for \ptot, \pt and
$x$-Feynman $\xF=2\,p_Z^*/\sqsnn$, where $p_Z^*$ is the longitudinal \pbar
momentum in the proton-nucleon center-of-mass system. These average values are obtained from
simulation, to avoid biases from reconstruction effects and given the
good agreement with data observed for the simulated kinematic spectra.
}
\label{tab:results}
\end{table}

\begin{small}  

\begin{tabular}{rrcrccrlrlrl}
\hline 
\multicolumn{2}{c}{\rule{0pt}{5ex} \ptot range} & $p_{\mathrm{\scriptscriptstyle T}}$ range & \multicolumn{1}{c}{$\langle\ptot\rangle$} &  $\langle p_{\mathrm{\scriptscriptstyle  T}}\rangle$ & $\langle \xF\rangle$ & \multicolumn{2}{c}{\dsigmaOndpdpt} & \multicolumn{2}{c}{$\delta_{\text{uncorr}}$} & \multicolumn{2}{c}{$\delta_{\text{corr}}$} \\ 
\multicolumn{2}{c}{\rule{0pt}{4ex} \small{[\gevc]}} &\small{[\gevc]} &\small{[\gevc]} &  \small{[\gevc]} & & \multicolumn{2}{c}{\small{$\left[\frac{\mubarn\,c^2}{\text{GeV}^2}\right]$}}  &  \multicolumn{2}{c}{\small{$\left[\frac{\mubarn\,c^2}{\text{GeV}^2}\right]$}}  & \multicolumn{2}{c}{\small{$\left[\frac{\mubarn\,c^2}{\text{GeV}^2}\right]$}}     \\  \hline 
  \\
 12.0 --~ &\!\!\!\!\!\!\!\!  14.0 &  0.6 --  0.7 &  12.99 & 0.62 & $ -0.050 $&$    324$ &$ $ &$      7$ &$ $ &$     26$ &$ $\\
 12.0 --~ &\!\!\!\!\!\!\!\!  14.0 &  0.7 --  0.8 &  12.99 & 0.75 & $ -0.057 $&$    241$ &$ $ &$     27$ &$ $ &$     19$ &$ $\\
 12.0 --~ &\!\!\!\!\!\!\!\!  14.0 &  0.8 --  0.9 &  12.99 & 0.85 & $ -0.063 $&$    188$ &$ $ &$     22$ &$ $ &$     15$ &$ $\\
 12.0 --~ &\!\!\!\!\!\!\!\!  14.0 &  0.9 --  1.1 &  12.99 & 0.97 & $ -0.073 $&$    122$ &$ $ &$     15$ &$ $ &$     10$ &$ $\\
 12.0 --~ &\!\!\!\!\!\!\!\!  14.0 &  1.1 --  1.2 &  12.99 & 1.12 & $ -0.085 $&$     80$ &$ $ &$     10$ &$ $ &$      5$ &$ $\\
 12.0 --~ &\!\!\!\!\!\!\!\!  14.0 &  1.2 --  1.5 &  12.99 & 1.32 & $ -0.106 $&$     38$ &$ \!\!\!\!\!\!\!.5$ &$      2$ &$ \!\!\!\!\!\!\!.7$ &$      2$ &$ \!\!\!\!\!\!\!.6$\\
 12.0 --~ &\!\!\!\!\!\!\!\!  14.0 &  1.5 --  2.0 &  12.99 & 1.67 & $ -0.149 $&$      8$ &$ \!\!\!\!\!\!\!.7$ &$      0$ &$ \!\!\!\!\!\!\!.7$ &$      0$ &$ \!\!\!\!\!\!\!.6$\\
 12.0 --~ &\!\!\!\!\!\!\!\!  14.0 &  2.0 --  2.8 &  12.99 & 2.21 & $ -0.236 $&$      0$ &$ \!\!\!\!\!\!\!.77$ &$      0$ &$ \!\!\!\!\!\!\!.11$ &$      0$ &$ \!\!\!\!\!\!\!.05$\\
 14.0 --~ &\!\!\!\!\!\!\!\!  16.2 &  0.6 --  0.7 &  15.09 & 0.62 & $ -0.042 $&$    312$ &$ $ &$      7$ &$ $ &$     25$ &$ $\\
 14.0 --~ &\!\!\!\!\!\!\!\!  16.2 &  0.7 --  0.8 &  15.09 & 0.75 & $ -0.048 $&$    245$ &$ $ &$      7$ &$ $ &$     20$ &$ $\\
 14.0 --~ &\!\!\!\!\!\!\!\!  16.2 &  0.8 --  0.9 &  15.09 & 0.85 & $ -0.054 $&$    195$ &$ \!\!\!\!\!\!\!.1$ &$      4$ &$ \!\!\!\!\!\!\!.9$ &$     15$ &$ \!\!\!\!\!\!\!.4$\\
 14.0 --~ &\!\!\!\!\!\!\!\!  16.2 &  0.9 --  1.1 &  15.09 & 0.97 & $ -0.062 $&$    135$ &$ \!\!\!\!\!\!\!.2$ &$      3$ &$ \!\!\!\!\!\!\!.4$ &$     10$ &$ \!\!\!\!\!\!\!.6$\\
 14.0 --~ &\!\!\!\!\!\!\!\!  16.2 &  1.1 --  1.2 &  15.09 & 1.12 & $ -0.073 $&$     80$ &$ \!\!\!\!\!\!\!.9$ &$      3$ &$ \!\!\!\!\!\!\!.1$ &$      5$ &$ \!\!\!\!\!\!\!.4$\\
 14.0 --~ &\!\!\!\!\!\!\!\!  16.2 &  1.2 --  1.5 &  15.09 & 1.32 & $ -0.091 $&$     40$ &$ \!\!\!\!\!\!\!.0$ &$      1$ &$ \!\!\!\!\!\!\!.3$ &$      2$ &$ \!\!\!\!\!\!\!.6$\\
 14.0 --~ &\!\!\!\!\!\!\!\!  16.2 &  1.5 --  2.0 &  15.09 & 1.67 & $ -0.128 $&$      9$ &$ \!\!\!\!\!\!\!.33$ &$      0$ &$ \!\!\!\!\!\!\!.39$ &$      0$ &$ \!\!\!\!\!\!\!.62$\\
 14.0 --~ &\!\!\!\!\!\!\!\!  16.2 &  2.0 --  2.8 &  15.09 & 2.21 & $ -0.202 $&$      1$ &$ \!\!\!\!\!\!\!.10$ &$      0$ &$ \!\!\!\!\!\!\!.11$ &$      0$ &$ \!\!\!\!\!\!\!.07$\\
\end{tabular}

\begin{tabular}{rrcrccrlrlrl}
\hline
\multicolumn{2}{c}{\rule{0pt}{5ex} \ptot range} & $p_{\mathrm{\scriptscriptstyle T}}$ range & \multicolumn{1}{c}{$\langle\ptot\rangle$} &  $\langle p_{\mathrm{\scriptscriptstyle  T}}\rangle$ & $\langle \xF\rangle$ & \multicolumn{2}{c}{\dsigmaOndpdpt} & \multicolumn{2}{c}{$\delta_{\text{uncorr}}$} & \multicolumn{2}{c}{$\delta_{\text{corr}}$} \\ 
\multicolumn{2}{c}{\rule{0pt}{4ex} \small{[\gevc]}} &\small{[\gevc]} &\small{[\gevc]} &  \small{[\gevc]} & & \multicolumn{2}{c}{\small{$\left[\frac{\mubarn\,c^2}{\text{GeV}^2}\right]$}}  &  \multicolumn{2}{c}{\small{$\left[\frac{\mubarn\,c^2}{\text{GeV}^2}\right]$}}  & \multicolumn{2}{c}{\small{$\left[\frac{\mubarn\,c^2}{\text{GeV}^2}\right]$}}     \\ \hline 
  \\
 16.2 --~ &\!\!\!\!\!\!\!\!  18.7 &  0.6 --  0.7 &  17.43 & 0.62 & $ -0.036 $&$    281$ &$ $ &$     10$ &$ $ &$     22$ &$ $\\
 16.2 --~ &\!\!\!\!\!\!\!\!  18.7 &  0.7 --  0.8 &  17.43 & 0.75 & $ -0.041 $&$    234$ &$ $ &$      6$ &$ $ &$     19$ &$ $\\
 16.2 --~ &\!\!\!\!\!\!\!\!  18.7 &  0.8 --  0.9 &  17.43 & 0.85 & $ -0.046 $&$    190$ &$ \!\!\!\!\!\!\!.2$ &$      4$ &$ \!\!\!\!\!\!\!.7$ &$     15$ &$ \!\!\!\!\!\!\!.1$\\
 16.2 --~ &\!\!\!\!\!\!\!\!  18.7 &  0.9 --  1.1 &  17.43 & 0.97 & $ -0.053 $&$    133$ &$ \!\!\!\!\!\!\!.5$ &$      3$ &$ \!\!\!\!\!\!\!.3$ &$     10$ &$ \!\!\!\!\!\!\!.6$\\
 16.2 --~ &\!\!\!\!\!\!\!\!  18.7 &  1.1 --  1.2 &  17.43 & 1.12 & $ -0.062 $&$     81$ &$ \!\!\!\!\!\!\!.0$ &$      2$ &$ \!\!\!\!\!\!\!.2$ &$      5$ &$ \!\!\!\!\!\!\!.4$\\
 16.2 --~ &\!\!\!\!\!\!\!\!  18.7 &  1.2 --  1.5 &  17.43 & 1.32 & $ -0.078 $&$     39$ &$ \!\!\!\!\!\!\!.2$ &$      1$ &$ \!\!\!\!\!\!\!.1$ &$      2$ &$ \!\!\!\!\!\!\!.6$\\
 16.2 --~ &\!\!\!\!\!\!\!\!  18.7 &  1.5 --  2.0 &  17.43 & 1.68 & $ -0.110 $&$     10$ &$ \!\!\!\!\!\!\!.44$ &$      0$ &$ \!\!\!\!\!\!\!.40$ &$      0$ &$ \!\!\!\!\!\!\!.69$\\
 16.2 --~ &\!\!\!\!\!\!\!\!  18.7 &  2.0 --  2.8 &  17.43 & 2.21 & $ -0.174 $&$      1$ &$ \!\!\!\!\!\!\!.03$ &$      0$ &$ \!\!\!\!\!\!\!.09$ &$      0$ &$ \!\!\!\!\!\!\!.07$\\
 18.7 --~ &\!\!\!\!\!\!\!\!  21.4 &  0.6 --  0.7 &  20.03 & 0.62 & $ -0.031 $&$    277$ &$ $ &$     19$ &$ $ &$     22$ &$ $\\
 18.7 --~ &\!\!\!\!\!\!\!\!  21.4 &  0.7 --  0.8 &  20.03 & 0.75 & $ -0.035 $&$    221$ &$ $ &$      5$ &$ $ &$     18$ &$ $\\
 18.7 --~ &\!\!\!\!\!\!\!\!  21.4 &  0.8 --  0.9 &  20.03 & 0.85 & $ -0.039 $&$    179$ &$ \!\!\!\!\!\!\!.1$ &$      4$ &$ \!\!\!\!\!\!\!.5$ &$     14$ &$ \!\!\!\!\!\!\!.2$\\
 18.7 --~ &\!\!\!\!\!\!\!\!  21.4 &  0.9 --  1.1 &  20.03 & 0.97 & $ -0.045 $&$    128$ &$ \!\!\!\!\!\!\!.3$ &$      3$ &$ \!\!\!\!\!\!\!.2$ &$     10$ &$ \!\!\!\!\!\!\!.2$\\
 18.7 --~ &\!\!\!\!\!\!\!\!  21.4 &  1.1 --  1.2 &  20.03 & 1.12 & $ -0.054 $&$     82$ &$ \!\!\!\!\!\!\!.2$ &$      2$ &$ \!\!\!\!\!\!\!.2$ &$      5$ &$ \!\!\!\!\!\!\!.5$\\
 18.7 --~ &\!\!\!\!\!\!\!\!  21.4 &  1.2 --  1.5 &  20.03 & 1.32 & $ -0.067 $&$     40$ &$ \!\!\!\!\!\!\!.1$ &$      1$ &$ \!\!\!\!\!\!\!.1$ &$      2$ &$ \!\!\!\!\!\!\!.7$\\
 18.7 --~ &\!\!\!\!\!\!\!\!  21.4 &  1.5 --  2.0 &  20.03 & 1.68 & $ -0.095 $&$     10$ &$ \!\!\!\!\!\!\!.44$ &$      0$ &$ \!\!\!\!\!\!\!.39$ &$      0$ &$ \!\!\!\!\!\!\!.69$\\
 18.7 --~ &\!\!\!\!\!\!\!\!  21.4 &  2.0 --  2.8 &  20.03 & 2.22 & $ -0.151 $&$      1$ &$ \!\!\!\!\!\!\!.16$ &$      0$ &$ \!\!\!\!\!\!\!.08$ &$      0$ &$ \!\!\!\!\!\!\!.07$\\
 21.4 --~ &\!\!\!\!\!\!\!\!  24.4 &  0.6 --  0.7 &  22.88 & 0.62 & $ -0.026 $&$    278$ &$ $ &$      6$ &$ $ &$     22$ &$ $\\
 21.4 --~ &\!\!\!\!\!\!\!\!  24.4 &  0.7 --  0.8 &  22.88 & 0.75 & $ -0.030 $&$    213$ &$ $ &$      5$ &$ $ &$     17$ &$ $\\
 21.4 --~ &\!\!\!\!\!\!\!\!  24.4 &  0.8 --  0.9 &  22.88 & 0.85 & $ -0.034 $&$    167$ &$ \!\!\!\!\!\!\!.2$ &$      4$ &$ \!\!\!\!\!\!\!.2$ &$     13$ &$ \!\!\!\!\!\!\!.3$\\
 21.4 --~ &\!\!\!\!\!\!\!\!  24.4 &  0.9 --  1.1 &  22.88 & 0.97 & $ -0.039 $&$    119$ &$ \!\!\!\!\!\!\!.5$ &$      3$ &$ \!\!\!\!\!\!\!.0$ &$      9$ &$ \!\!\!\!\!\!\!.5$\\
 21.4 --~ &\!\!\!\!\!\!\!\!  24.4 &  1.1 --  1.2 &  22.88 & 1.12 & $ -0.046 $&$     78$ &$ \!\!\!\!\!\!\!.0$ &$      2$ &$ \!\!\!\!\!\!\!.1$ &$      5$ &$ \!\!\!\!\!\!\!.3$\\
 21.4 --~ &\!\!\!\!\!\!\!\!  24.4 &  1.2 --  1.5 &  22.88 & 1.32 & $ -0.058 $&$     37$ &$ \!\!\!\!\!\!\!.7$ &$      1$ &$ \!\!\!\!\!\!\!.1$ &$      2$ &$ \!\!\!\!\!\!\!.6$\\
 21.4 --~ &\!\!\!\!\!\!\!\!  24.4 &  1.5 --  2.0 &  22.88 & 1.68 & $ -0.083 $&$     10$ &$ \!\!\!\!\!\!\!.38$ &$      0$ &$ \!\!\!\!\!\!\!.36$ &$      0$ &$ \!\!\!\!\!\!\!.68$\\
 21.4 --~ &\!\!\!\!\!\!\!\!  24.4 &  2.0 --  2.8 &  22.88 & 2.22 & $ -0.132 $&$      1$ &$ \!\!\!\!\!\!\!.19$ &$      0$ &$ \!\!\!\!\!\!\!.09$ &$      0$ &$ \!\!\!\!\!\!\!.08$\\
 24.4 --~ &\!\!\!\!\!\!\!\!  27.7 &  0.4 --  0.6 &  26.02 & 0.47 & $ -0.019 $&$    519$ &$ $ &$    185$ &$ $ &$     44$ &$ $\\
 24.4 --~ &\!\!\!\!\!\!\!\!  27.7 &  0.6 --  0.7 &  26.02 & 0.62 & $ -0.022 $&$    289$ &$ $ &$     13$ &$ $ &$     24$ &$ $\\
 24.4 --~ &\!\!\!\!\!\!\!\!  27.7 &  0.7 --  0.8 &  26.02 & 0.75 & $ -0.025 $&$    205$ &$ $ &$      5$ &$ $ &$     16$ &$ $\\
 24.4 --~ &\!\!\!\!\!\!\!\!  27.7 &  0.8 --  0.9 &  26.02 & 0.85 & $ -0.029 $&$    156$ &$ \!\!\!\!\!\!\!.2$ &$      3$ &$ \!\!\!\!\!\!\!.9$ &$     12$ &$ \!\!\!\!\!\!\!.4$\\
 24.4 --~ &\!\!\!\!\!\!\!\!  27.7 &  0.9 --  1.1 &  26.02 & 0.97 & $ -0.033 $&$    110$ &$ \!\!\!\!\!\!\!.6$ &$      2$ &$ \!\!\!\!\!\!\!.7$ &$      8$ &$ \!\!\!\!\!\!\!.8$\\
 24.4 --~ &\!\!\!\!\!\!\!\!  27.7 &  1.1 --  1.2 &  26.02 & 1.12 & $ -0.040 $&$     72$ &$ \!\!\!\!\!\!\!.8$ &$      1$ &$ \!\!\!\!\!\!\!.9$ &$      4$ &$ \!\!\!\!\!\!\!.9$\\
 24.4 --~ &\!\!\!\!\!\!\!\!  27.7 &  1.2 --  1.5 &  26.02 & 1.32 & $ -0.050 $&$     37$ &$ \!\!\!\!\!\!\!.0$ &$      1$ &$ \!\!\!\!\!\!\!.0$ &$      2$ &$ \!\!\!\!\!\!\!.5$\\
 24.4 --~ &\!\!\!\!\!\!\!\!  27.7 &  1.5 --  2.0 &  26.02 & 1.68 & $ -0.072 $&$      9$ &$ \!\!\!\!\!\!\!.94$ &$      0$ &$ \!\!\!\!\!\!\!.33$ &$      0$ &$ \!\!\!\!\!\!\!.67$\\
 24.4 --~ &\!\!\!\!\!\!\!\!  27.7 &  2.0 --  2.8 &  26.02 & 2.23 & $ -0.116 $&$      1$ &$ \!\!\!\!\!\!\!.29$ &$      0$ &$ \!\!\!\!\!\!\!.08$ &$      0$ &$ \!\!\!\!\!\!\!.08$\\
 27.7 --~ &\!\!\!\!\!\!\!\!  31.4 &  0.4 --  0.6 &  29.52 & 0.47 & $ -0.015 $&$    451$ &$ $ &$    116$ &$ $ &$     38$ &$ $\\
 27.7 --~ &\!\!\!\!\!\!\!\!  31.4 &  0.6 --  0.7 &  29.52 & 0.62 & $ -0.018 $&$    318$ &$ $ &$     45$ &$ $ &$     27$ &$ $\\
 27.7 --~ &\!\!\!\!\!\!\!\!  31.4 &  0.7 --  0.8 &  29.52 & 0.75 & $ -0.021 $&$    219$ &$ $ &$      5$ &$ $ &$     18$ &$ $\\
 27.7 --~ &\!\!\!\!\!\!\!\!  31.4 &  0.8 --  0.9 &  29.52 & 0.85 & $ -0.024 $&$    152$ &$ \!\!\!\!\!\!\!.2$ &$      3$ &$ \!\!\!\!\!\!\!.8$ &$     12$ &$ \!\!\!\!\!\!\!.2$\\
 27.7 --~ &\!\!\!\!\!\!\!\!  31.4 &  0.9 --  1.1 &  29.52 & 0.97 & $ -0.028 $&$    103$ &$ \!\!\!\!\!\!\!.5$ &$      2$ &$ \!\!\!\!\!\!\!.6$ &$      8$ &$ \!\!\!\!\!\!\!.2$\\
 27.7 --~ &\!\!\!\!\!\!\!\!  31.4 &  1.1 --  1.2 &  29.52 & 1.12 & $ -0.034 $&$     67$ &$ \!\!\!\!\!\!\!.8$ &$      1$ &$ \!\!\!\!\!\!\!.8$ &$      4$ &$ \!\!\!\!\!\!\!.6$\\
 27.7 --~ &\!\!\!\!\!\!\!\!  31.4 &  1.2 --  1.5 &  29.52 & 1.33 & $ -0.043 $&$     33$ &$ \!\!\!\!\!\!\!.9$ &$      1$ &$ \!\!\!\!\!\!\!.0$ &$      2$ &$ \!\!\!\!\!\!\!.3$\\
 27.7 --~ &\!\!\!\!\!\!\!\!  31.4 &  1.5 --  2.0 &  29.52 & 1.68 & $ -0.062 $&$      9$ &$ \!\!\!\!\!\!\!.89$ &$      0$ &$ \!\!\!\!\!\!\!.32$ &$      0$ &$ \!\!\!\!\!\!\!.67$\\
 27.7 --~ &\!\!\!\!\!\!\!\!  31.4 &  2.0 --  2.8 &  29.52 & 2.23 & $ -0.101 $&$      1$ &$ \!\!\!\!\!\!\!.28$ &$      0$ &$ \!\!\!\!\!\!\!.08$ &$      0$ &$ \!\!\!\!\!\!\!.08$\\
\end{tabular}

\begin{tabular}{rrcrccrlrlrl}
\hline
\multicolumn{2}{c}{\rule{0pt}{5ex} \ptot range} & $p_{\mathrm{\scriptscriptstyle T}}$ range & \multicolumn{1}{c}{$\langle\ptot\rangle$} &  $\langle p_{\mathrm{\scriptscriptstyle  T}}\rangle$ & $\langle \xF\rangle$ & \multicolumn{2}{c}{\dsigmaOndpdpt} & \multicolumn{2}{c}{$\delta_{\text{uncorr}}$} & \multicolumn{2}{c}{$\delta_{\text{corr}}$} \\ 
\multicolumn{2}{c}{\rule{0pt}{4ex} \small{[\gevc]}} &\small{[\gevc]} &\small{[\gevc]} &  \small{[\gevc]} & & \multicolumn{2}{c}{\small{$\left[\frac{\mubarn\,c^2}{\text{GeV}^2}\right]$}}  &  \multicolumn{2}{c}{\small{$\left[\frac{\mubarn\,c^2}{\text{GeV}^2}\right]$}}  & \multicolumn{2}{c}{\small{$\left[\frac{\mubarn\,c^2}{\text{GeV}^2}\right]$}}     \\ \hline 
  \\
 31.4 --~ &\!\!\!\!\!\!\!\!  35.5 &  0.4 --  0.6 &  33.41 & 0.47 & $ -0.012 $&$    339$ &$ $ &$     75$ &$ $ &$     31$ &$ $\\
 31.4 --~ &\!\!\!\!\!\!\!\!  35.5 &  0.6 --  0.7 &  33.41 & 0.62 & $ -0.015 $&$    274$ &$ $ &$     54$ &$ $ &$     23$ &$ $\\
 31.4 --~ &\!\!\!\!\!\!\!\!  35.5 &  0.7 --  0.8 &  33.41 & 0.75 & $ -0.018 $&$    195$ &$ $ &$     15$ &$ $ &$     16$ &$ $\\
 31.4 --~ &\!\!\!\!\!\!\!\!  35.5 &  0.8 --  0.9 &  33.41 & 0.85 & $ -0.020 $&$    136$ &$ \!\!\!\!\!\!\!.4$ &$      3$ &$ \!\!\!\!\!\!\!.7$ &$     11$ &$ \!\!\!\!\!\!\!.0$\\
 31.4 --~ &\!\!\!\!\!\!\!\!  35.5 &  0.9 --  1.1 &  33.41 & 0.97 & $ -0.024 $&$     95$ &$ \!\!\!\!\!\!\!.0$ &$      2$ &$ \!\!\!\!\!\!\!.4$ &$      7$ &$ \!\!\!\!\!\!\!.6$\\
 31.4 --~ &\!\!\!\!\!\!\!\!  35.5 &  1.1 --  1.2 &  33.41 & 1.12 & $ -0.029 $&$     62$ &$ \!\!\!\!\!\!\!.5$ &$      1$ &$ \!\!\!\!\!\!\!.7$ &$      4$ &$ \!\!\!\!\!\!\!.2$\\
 31.4 --~ &\!\!\!\!\!\!\!\!  35.5 &  1.2 --  1.5 &  33.41 & 1.33 & $ -0.037 $&$     32$ &$ \!\!\!\!\!\!\!.0$ &$      0$ &$ \!\!\!\!\!\!\!.9$ &$      2$ &$ \!\!\!\!\!\!\!.2$\\
 31.4 --~ &\!\!\!\!\!\!\!\!  35.5 &  1.5 --  2.0 &  33.41 & 1.68 & $ -0.054 $&$      9$ &$ \!\!\!\!\!\!\!.58$ &$      0$ &$ \!\!\!\!\!\!\!.31$ &$      0$ &$ \!\!\!\!\!\!\!.64$\\
 31.4 --~ &\!\!\!\!\!\!\!\!  35.5 &  2.0 --  2.8 &  33.41 & 2.23 & $ -0.088 $&$      1$ &$ \!\!\!\!\!\!\!.40$ &$      0$ &$ \!\!\!\!\!\!\!.07$ &$      0$ &$ \!\!\!\!\!\!\!.09$\\
 35.5 --~ &\!\!\!\!\!\!\!\!  40.0 &  0.4 --  0.6 &  37.71 & 0.47 & $ -0.010 $&$    267$ &$ $ &$     39$ &$ $ &$     25$ &$ $\\
 35.5 --~ &\!\!\!\!\!\!\!\!  40.0 &  0.6 --  0.7 &  37.71 & 0.62 & $ -0.012 $&$    240$ &$ $ &$     11$ &$ $ &$     21$ &$ $\\
 35.5 --~ &\!\!\!\!\!\!\!\!  40.0 &  0.7 --  0.8 &  37.71 & 0.75 & $ -0.015 $&$    177$ &$ $ &$     13$ &$ $ &$     16$ &$ $\\
 35.5 --~ &\!\!\!\!\!\!\!\!  40.0 &  0.8 --  0.9 &  37.71 & 0.85 & $ -0.017 $&$    125$ &$ \!\!\!\!\!\!\!.1$ &$      3$ &$ \!\!\!\!\!\!\!.3$ &$     10$ &$ \!\!\!\!\!\!\!.9$\\
 35.5 --~ &\!\!\!\!\!\!\!\!  40.0 &  0.9 --  1.1 &  37.71 & 0.97 & $ -0.020 $&$     86$ &$ \!\!\!\!\!\!\!.9$ &$      2$ &$ \!\!\!\!\!\!\!.2$ &$      6$ &$ \!\!\!\!\!\!\!.0$\\
 35.5 --~ &\!\!\!\!\!\!\!\!  40.0 &  1.1 --  1.2 &  37.71 & 1.12 & $ -0.024 $&$     57$ &$ \!\!\!\!\!\!\!.7$ &$      1$ &$ \!\!\!\!\!\!\!.6$ &$      3$ &$ \!\!\!\!\!\!\!.9$\\
 35.5 --~ &\!\!\!\!\!\!\!\!  40.0 &  1.2 --  1.5 &  37.71 & 1.33 & $ -0.032 $&$     30$ &$ \!\!\!\!\!\!\!.6$ &$      0$ &$ \!\!\!\!\!\!\!.8$ &$      2$ &$ \!\!\!\!\!\!\!.1$\\
 35.5 --~ &\!\!\!\!\!\!\!\!  40.0 &  1.5 --  2.0 &  37.71 & 1.68 & $ -0.047 $&$      9$ &$ \!\!\!\!\!\!\!.11$ &$      0$ &$ \!\!\!\!\!\!\!.29$ &$      0$ &$ \!\!\!\!\!\!\!.61$\\
 35.5 --~ &\!\!\!\!\!\!\!\!  40.0 &  2.0 --  2.8 &  37.71 & 2.23 & $ -0.077 $&$      1$ &$ \!\!\!\!\!\!\!.34$ &$      0$ &$ \!\!\!\!\!\!\!.07$ &$      0$ &$ \!\!\!\!\!\!\!.09$\\
 35.5 --~ &\!\!\!\!\!\!\!\!  40.0 &  2.8 --  4.0 &  37.71 & 3.06 & $ -0.139 $&$      0$ &$ \!\!\!\!\!\!\!.065$ &$      0$ &$ \!\!\!\!\!\!\!.012$ &$      0$ &$ \!\!\!\!\!\!\!.004$\\
 40.0 --~ &\!\!\!\!\!\!\!\!  45.0 &  0.6 --  0.7 &  42.46 & 0.62 & $ -0.009 $&$    192$ &$ $ &$     12$ &$ $ &$     17$ &$ $\\
 40.0 --~ &\!\!\!\!\!\!\!\!  45.0 &  0.7 --  0.8 &  42.46 & 0.75 & $ -0.012 $&$    148$ &$ $ &$      5$ &$ $ &$     13$ &$ $\\
 40.0 --~ &\!\!\!\!\!\!\!\!  45.0 &  0.8 --  0.9 &  42.46 & 0.85 & $ -0.014 $&$    110$ &$ $ &$      7$ &$ $ &$     10$ &$ $\\
 40.0 --~ &\!\!\!\!\!\!\!\!  45.0 &  0.9 --  1.1 &  42.46 & 0.97 & $ -0.016 $&$     79$ &$ \!\!\!\!\!\!\!.4$ &$      2$ &$ \!\!\!\!\!\!\!.1$ &$      6$ &$ \!\!\!\!\!\!\!.9$\\
 40.0 --~ &\!\!\!\!\!\!\!\!  45.0 &  1.1 --  1.2 &  42.46 & 1.12 & $ -0.020 $&$     49$ &$ \!\!\!\!\!\!\!.8$ &$      1$ &$ \!\!\!\!\!\!\!.4$ &$      3$ &$ \!\!\!\!\!\!\!.4$\\
 40.0 --~ &\!\!\!\!\!\!\!\!  45.0 &  1.2 --  1.5 &  42.46 & 1.33 & $ -0.027 $&$     27$ &$ \!\!\!\!\!\!\!.4$ &$      0$ &$ \!\!\!\!\!\!\!.7$ &$      1$ &$ \!\!\!\!\!\!\!.8$\\
 40.0 --~ &\!\!\!\!\!\!\!\!  45.0 &  1.5 --  2.0 &  42.46 & 1.69 & $ -0.040 $&$      8$ &$ \!\!\!\!\!\!\!.79$ &$      0$ &$ \!\!\!\!\!\!\!.27$ &$      0$ &$ \!\!\!\!\!\!\!.59$\\
 40.0 --~ &\!\!\!\!\!\!\!\!  45.0 &  2.0 --  2.8 &  42.46 & 2.24 & $ -0.067 $&$      1$ &$ \!\!\!\!\!\!\!.26$ &$      0$ &$ \!\!\!\!\!\!\!.06$ &$      0$ &$ \!\!\!\!\!\!\!.08$\\
 40.0 --~ &\!\!\!\!\!\!\!\!  45.0 &  2.8 --  4.0 &  42.46 & 3.08 & $ -0.124 $&$      0$ &$ \!\!\!\!\!\!\!.059$ &$      0$ &$ \!\!\!\!\!\!\!.010$ &$      0$ &$ \!\!\!\!\!\!\!.004$\\
 45.0 --~ &\!\!\!\!\!\!\!\!  50.5 &  0.6 --  0.7 &  47.70 & 0.62 & $ -0.007 $&$    151$ &$ \!\!\!\!\!\!\!.4$ &$      3$ &$ \!\!\!\!\!\!\!.9$ &$     14$ &$ \!\!\!\!\!\!\!.0$\\
 45.0 --~ &\!\!\!\!\!\!\!\!  50.5 &  0.7 --  0.8 &  47.70 & 0.75 & $ -0.009 $&$    130$ &$ \!\!\!\!\!\!\!.0$ &$      3$ &$ \!\!\!\!\!\!\!.4$ &$     11$ &$ \!\!\!\!\!\!\!.5$\\
 45.0 --~ &\!\!\!\!\!\!\!\!  50.5 &  0.8 --  0.9 &  47.70 & 0.85 & $ -0.011 $&$    100$ &$ \!\!\!\!\!\!\!.8$ &$      3$ &$ \!\!\!\!\!\!\!.4$ &$      9$ &$ \!\!\!\!\!\!\!.0$\\
 45.0 --~ &\!\!\!\!\!\!\!\!  50.5 &  0.9 --  1.1 &  47.70 & 0.97 & $ -0.013 $&$     70$ &$ \!\!\!\!\!\!\!.8$ &$      1$ &$ \!\!\!\!\!\!\!.9$ &$      6$ &$ \!\!\!\!\!\!\!.3$\\
 45.0 --~ &\!\!\!\!\!\!\!\!  50.5 &  1.1 --  1.2 &  47.70 & 1.12 & $ -0.016 $&$     45$ &$ \!\!\!\!\!\!\!.5$ &$      2$ &$ \!\!\!\!\!\!\!.4$ &$      3$ &$ \!\!\!\!\!\!\!.2$\\
 45.0 --~ &\!\!\!\!\!\!\!\!  50.5 &  1.2 --  1.5 &  47.70 & 1.33 & $ -0.022 $&$     23$ &$ \!\!\!\!\!\!\!.7$ &$      0$ &$ \!\!\!\!\!\!\!.6$ &$      1$ &$ \!\!\!\!\!\!\!.6$\\
 45.0 --~ &\!\!\!\!\!\!\!\!  50.5 &  1.5 --  2.0 &  47.70 & 1.69 & $ -0.034 $&$      8$ &$ \!\!\!\!\!\!\!.38$ &$      0$ &$ \!\!\!\!\!\!\!.26$ &$      0$ &$ \!\!\!\!\!\!\!.56$\\
 45.0 --~ &\!\!\!\!\!\!\!\!  50.5 &  2.0 --  2.8 &  47.70 & 2.24 & $ -0.058 $&$      1$ &$ \!\!\!\!\!\!\!.29$ &$      0$ &$ \!\!\!\!\!\!\!.06$ &$      0$ &$ \!\!\!\!\!\!\!.09$\\
 45.0 --~ &\!\!\!\!\!\!\!\!  50.5 &  2.8 --  4.0 &  47.70 & 3.09 & $ -0.109 $&$      0$ &$ \!\!\!\!\!\!\!.059$ &$      0$ &$ \!\!\!\!\!\!\!.009$ &$      0$ &$ \!\!\!\!\!\!\!.004$\\
 50.5 --~ &\!\!\!\!\!\!\!\!  56.7 &  0.7 --  0.8 &  53.54 & 0.75 & $ -0.006 $&$    109$ &$ \!\!\!\!\!\!\!.2$ &$      3$ &$ \!\!\!\!\!\!\!.1$ &$      9$ &$ \!\!\!\!\!\!\!.9$\\
 50.5 --~ &\!\!\!\!\!\!\!\!  56.7 &  0.8 --  0.9 &  53.54 & 0.85 & $ -0.008 $&$     86$ &$ \!\!\!\!\!\!\!.6$ &$      2$ &$ \!\!\!\!\!\!\!.4$ &$      7$ &$ \!\!\!\!\!\!\!.6$\\
 50.5 --~ &\!\!\!\!\!\!\!\!  56.7 &  0.9 --  1.1 &  53.54 & 0.97 & $ -0.010 $&$     65$ &$ \!\!\!\!\!\!\!.8$ &$      1$ &$ \!\!\!\!\!\!\!.8$ &$      5$ &$ \!\!\!\!\!\!\!.8$\\
 50.5 --~ &\!\!\!\!\!\!\!\!  56.7 &  1.1 --  1.2 &  53.54 & 1.12 & $ -0.013 $&$     40$ &$ \!\!\!\!\!\!\!.3$ &$      1$ &$ \!\!\!\!\!\!\!.2$ &$      3$ &$ \!\!\!\!\!\!\!.5$\\
 50.5 --~ &\!\!\!\!\!\!\!\!  56.7 &  1.2 --  1.5 &  53.54 & 1.33 & $ -0.018 $&$     21$ &$ \!\!\!\!\!\!\!.0$ &$      0$ &$ \!\!\!\!\!\!\!.7$ &$      1$ &$ \!\!\!\!\!\!\!.5$\\
 50.5 --~ &\!\!\!\!\!\!\!\!  56.7 &  1.5 --  2.0 &  53.54 & 1.69 & $ -0.029 $&$      7$ &$ \!\!\!\!\!\!\!.56$ &$      0$ &$ \!\!\!\!\!\!\!.23$ &$      0$ &$ \!\!\!\!\!\!\!.51$\\
 50.5 --~ &\!\!\!\!\!\!\!\!  56.7 &  2.0 --  2.8 &  53.54 & 2.24 & $ -0.051 $&$      1$ &$ \!\!\!\!\!\!\!.18$ &$      0$ &$ \!\!\!\!\!\!\!.05$ &$      0$ &$ \!\!\!\!\!\!\!.08$\\
 50.5 --~ &\!\!\!\!\!\!\!\!  56.7 &  2.8 --  4.0 &  53.54 & 3.09 & $ -0.096 $&$      0$ &$ \!\!\!\!\!\!\!.070$ &$      0$ &$ \!\!\!\!\!\!\!.010$ &$      0$ &$ \!\!\!\!\!\!\!.005$\\
\end{tabular}

\begin{tabular}{rrcrccrlrlrl}
\hline
\multicolumn{2}{c}{\rule{0pt}{5ex} \ptot range} & $p_{\mathrm{\scriptscriptstyle T}}$ range & \multicolumn{1}{c}{$\langle\ptot\rangle$} &  $\langle p_{\mathrm{\scriptscriptstyle  T}}\rangle$ & $\langle \xF\rangle$ & \multicolumn{2}{c}{\dsigmaOndpdpt} & \multicolumn{2}{c}{$\delta_{\text{uncorr}}$} & \multicolumn{2}{c}{$\delta_{\text{corr}}$} \\ 
\multicolumn{2}{c}{\rule{0pt}{4ex} \small{[\gevc]}} &\small{[\gevc]} &\small{[\gevc]} &  \small{[\gevc]} & & \multicolumn{2}{c}{\small{$\left[\frac{\mubarn\,c^2}{\text{GeV}^2}\right]$}}  &  \multicolumn{2}{c}{\small{$\left[\frac{\mubarn\,c^2}{\text{GeV}^2}\right]$}}  & \multicolumn{2}{c}{\small{$\left[\frac{\mubarn\,c^2}{\text{GeV}^2}\right]$}}     \\ \hline 
  \\
 56.7 --~ &\!\!\!\!\!\!\!\!  63.5 &  0.8 --  0.9 &  60.04 & 0.85 & $ -0.005 $&$     74$ &$ \!\!\!\!\!\!\!.1$ &$      2$ &$ \!\!\!\!\!\!\!.2$ &$      6$ &$ \!\!\!\!\!\!\!.6$\\
 56.7 --~ &\!\!\!\!\!\!\!\!  63.5 &  0.9 --  1.1 &  60.04 & 0.97 & $ -0.007 $&$     57$ &$ \!\!\!\!\!\!\!.8$ &$      1$ &$ \!\!\!\!\!\!\!.6$ &$      5$ &$ \!\!\!\!\!\!\!.1$\\
 56.7 --~ &\!\!\!\!\!\!\!\!  63.5 &  1.1 --  1.2 &  60.04 & 1.12 & $ -0.010 $&$     37$ &$ \!\!\!\!\!\!\!.0$ &$      1$ &$ \!\!\!\!\!\!\!.1$ &$      3$ &$ \!\!\!\!\!\!\!.3$\\
 56.7 --~ &\!\!\!\!\!\!\!\!  63.5 &  1.2 --  1.5 &  60.04 & 1.33 & $ -0.014 $&$     18$ &$ \!\!\!\!\!\!\!.7$ &$      0$ &$ \!\!\!\!\!\!\!.5$ &$      1$ &$ \!\!\!\!\!\!\!.6$\\
 56.7 --~ &\!\!\!\!\!\!\!\!  63.5 &  1.5 --  2.0 &  60.04 & 1.69 & $ -0.024 $&$      6$ &$ \!\!\!\!\!\!\!.79$ &$      0$ &$ \!\!\!\!\!\!\!.21$ &$      0$ &$ \!\!\!\!\!\!\!.46$\\
 56.7 --~ &\!\!\!\!\!\!\!\!  63.5 &  2.0 --  2.8 &  60.04 & 2.24 & $ -0.043 $&$      1$ &$ \!\!\!\!\!\!\!.22$ &$      0$ &$ \!\!\!\!\!\!\!.06$ &$      0$ &$ \!\!\!\!\!\!\!.08$\\
 56.7 --~ &\!\!\!\!\!\!\!\!  63.5 &  2.8 --  4.0 &  60.04 & 3.09 & $ -0.083 $&$      0$ &$ \!\!\!\!\!\!\!.071$ &$      0$ &$ \!\!\!\!\!\!\!.010$ &$      0$ &$ \!\!\!\!\!\!\!.005$\\
 63.5 --~ &\!\!\!\!\!\!\!\!  71.0 &  0.8 --  0.9 &  67.18 & 0.85 & $ -0.002 $&$     64$ &$ \!\!\!\!\!\!\!.6$ &$      2$ &$ \!\!\!\!\!\!\!.4$ &$      6$ &$ \!\!\!\!\!\!\!.2$\\
 63.5 --~ &\!\!\!\!\!\!\!\!  71.0 &  0.9 --  1.1 &  67.18 & 0.97 & $ -0.004 $&$     51$ &$ \!\!\!\!\!\!\!.7$ &$      1$ &$ \!\!\!\!\!\!\!.5$ &$      4$ &$ \!\!\!\!\!\!\!.6$\\
 63.5 --~ &\!\!\!\!\!\!\!\!  71.0 &  1.1 --  1.2 &  67.18 & 1.12 & $ -0.007 $&$     35$ &$ \!\!\!\!\!\!\!.2$ &$      1$ &$ \!\!\!\!\!\!\!.1$ &$      3$ &$ \!\!\!\!\!\!\!.1$\\
 63.5 --~ &\!\!\!\!\!\!\!\!  71.0 &  1.2 --  1.5 &  67.18 & 1.33 & $ -0.011 $&$     17$ &$ \!\!\!\!\!\!\!.7$ &$      1$ &$ \!\!\!\!\!\!\!.0$ &$      1$ &$ \!\!\!\!\!\!\!.6$\\
 63.5 --~ &\!\!\!\!\!\!\!\!  71.0 &  1.5 --  2.0 &  67.18 & 1.69 & $ -0.019 $&$      6$ &$ \!\!\!\!\!\!\!.25$ &$      0$ &$ \!\!\!\!\!\!\!.20$ &$      0$ &$ \!\!\!\!\!\!\!.43$\\
 63.5 --~ &\!\!\!\!\!\!\!\!  71.0 &  2.0 --  2.8 &  67.18 & 2.24 & $ -0.037 $&$      1$ &$ \!\!\!\!\!\!\!.15$ &$      0$ &$ \!\!\!\!\!\!\!.05$ &$      0$ &$ \!\!\!\!\!\!\!.08$\\
 63.5 --~ &\!\!\!\!\!\!\!\!  71.0 &  2.8 --  4.0 &  67.18 & 3.09 & $ -0.072 $&$      0$ &$ \!\!\!\!\!\!\!.081$ &$      0$ &$ \!\!\!\!\!\!\!.012$ &$      0$ &$ \!\!\!\!\!\!\!.005$\\
 71.0 --~ &\!\!\!\!\!\!\!\!  79.3 &  0.9 --  1.1 &  75.07 & 0.97 & $ -0.001 $&$     44$ &$ \!\!\!\!\!\!\!.0$ &$      1$ &$ \!\!\!\!\!\!\!.6$ &$      4$ &$ \!\!\!\!\!\!\!.1$\\
 71.0 --~ &\!\!\!\!\!\!\!\!  79.3 &  1.1 --  1.2 &  75.07 & 1.12 & $ -0.004 $&$     29$ &$ \!\!\!\!\!\!\!.6$ &$      0$ &$ \!\!\!\!\!\!\!.9$ &$      2$ &$ \!\!\!\!\!\!\!.6$\\
 71.0 --~ &\!\!\!\!\!\!\!\!  79.3 &  1.2 --  1.5 &  75.07 & 1.33 & $ -0.007 $&$     16$ &$ \!\!\!\!\!\!\!.00$ &$      0$ &$ \!\!\!\!\!\!\!.48$ &$      1$ &$ \!\!\!\!\!\!\!.40$\\
 71.0 --~ &\!\!\!\!\!\!\!\!  79.3 &  1.5 --  2.0 &  75.07 & 1.69 & $ -0.015 $&$      5$ &$ \!\!\!\!\!\!\!.23$ &$      0$ &$ \!\!\!\!\!\!\!.17$ &$      0$ &$ \!\!\!\!\!\!\!.46$\\
 71.0 --~ &\!\!\!\!\!\!\!\!  79.3 &  2.0 --  2.8 &  75.07 & 2.24 & $ -0.030 $&$      1$ &$ \!\!\!\!\!\!\!.02$ &$      0$ &$ \!\!\!\!\!\!\!.05$ &$      0$ &$ \!\!\!\!\!\!\!.07$\\
 71.0 --~ &\!\!\!\!\!\!\!\!  79.3 &  2.8 --  4.0 &  75.07 & 3.10 & $ -0.063 $&$      0$ &$ \!\!\!\!\!\!\!.069$ &$      0$ &$ \!\!\!\!\!\!\!.009$ &$      0$ &$ \!\!\!\!\!\!\!.005$\\
 79.3 --~ &\!\!\!\!\!\!\!\!  88.5 &  1.1 --  1.2 &  83.81 & 1.12 & $ -0.001 $&$     25$ &$ \!\!\!\!\!\!\!.1$ &$      1$ &$ \!\!\!\!\!\!\!.1$ &$      2$ &$ \!\!\!\!\!\!\!.3$\\
 79.3 --~ &\!\!\!\!\!\!\!\!  88.5 &  1.2 --  1.5 &  83.81 & 1.33 & $ -0.004 $&$     14$ &$ \!\!\!\!\!\!\!.64$ &$      0$ &$ \!\!\!\!\!\!\!.46$ &$      1$ &$ \!\!\!\!\!\!\!.30$\\
 79.3 --~ &\!\!\!\!\!\!\!\!  88.5 &  1.5 --  2.0 &  83.81 & 1.69 & $ -0.011 $&$      4$ &$ \!\!\!\!\!\!\!.75$ &$      0$ &$ \!\!\!\!\!\!\!.16$ &$      0$ &$ \!\!\!\!\!\!\!.42$\\
 79.3 --~ &\!\!\!\!\!\!\!\!  88.5 &  2.0 --  2.8 &  83.81 & 2.25 & $ -0.025 $&$      0$ &$ \!\!\!\!\!\!\!.93$ &$      0$ &$ \!\!\!\!\!\!\!.04$ &$      0$ &$ \!\!\!\!\!\!\!.07$\\
 79.3 --~ &\!\!\!\!\!\!\!\!  88.5 &  2.8 --  4.0 &  83.81 & 3.11 & $ -0.054 $&$      0$ &$ \!\!\!\!\!\!\!.069$ &$      0$ &$ \!\!\!\!\!\!\!.008$ &$      0$ &$ \!\!\!\!\!\!\!.005$\\
 88.5 --~ &\!\!\!\!\!\!\!\!  98.7 &  1.2 --  1.5 &  93.50 & 1.33 & $ -0.001 $&$     13$ &$ \!\!\!\!\!\!\!.43$ &$      0$ &$ \!\!\!\!\!\!\!.49$ &$      1$ &$ \!\!\!\!\!\!\!.21$\\
 88.5 --~ &\!\!\!\!\!\!\!\!  98.7 &  1.5 --  2.0 &  93.50 & 1.69 & $ -0.007 $&$      4$ &$ \!\!\!\!\!\!\!.41$ &$      0$ &$ \!\!\!\!\!\!\!.46$ &$      0$ &$ \!\!\!\!\!\!\!.39$\\
 88.5 --~ &\!\!\!\!\!\!\!\!  98.7 &  2.0 --  2.8 &  93.50 & 2.25 & $ -0.019 $&$      0$ &$ \!\!\!\!\!\!\!.81$ &$      0$ &$ \!\!\!\!\!\!\!.04$ &$      0$ &$ \!\!\!\!\!\!\!.06$\\
 88.5 --~ &\!\!\!\!\!\!\!\!  98.7 &  2.8 --  4.0 &  93.50 & 3.11 & $ -0.046 $&$      0$ &$ \!\!\!\!\!\!\!.064$ &$      0$ &$ \!\!\!\!\!\!\!.011$ &$      0$ &$ \!\!\!\!\!\!\!.004$\\
 98.7 --~ &\!\!\!\!\!\!\!\! 110.0 &  1.2 --  1.5 & 104.23 & 1.33 & $ +0.003 $&$     10$ &$ \!\!\!\!\!\!\!.8$ &$      1$ &$ \!\!\!\!\!\!\!.5$ &$      1$ &$ \!\!\!\!\!\!\!.0$\\
 98.7 --~ &\!\!\!\!\!\!\!\! 110.0 &  1.5 --  2.0 & 104.23 & 1.69 & $ -0.003 $&$      3$ &$ \!\!\!\!\!\!\!.83$ &$      0$ &$ \!\!\!\!\!\!\!.69$ &$      0$ &$ \!\!\!\!\!\!\!.34$\\
 98.7 --~ &\!\!\!\!\!\!\!\! 110.0 &  2.0 --  2.8 & 104.23 & 2.25 & $ -0.014 $&$      0$ &$ \!\!\!\!\!\!\!.68$ &$      0$ &$ \!\!\!\!\!\!\!.07$ &$      0$ &$ \!\!\!\!\!\!\!.06$\\
 98.7 --~ &\!\!\!\!\!\!\!\! 110.0 &  2.8 --  4.0 & 104.23 & 3.12 & $ -0.038 $&$      0$ &$ \!\!\!\!\!\!\!.052$ &$      0$ &$ \!\!\!\!\!\!\!.008$ &$      0$ &$ \!\!\!\!\!\!\!.003$\\
\end{tabular}

\end{small}


\addcontentsline{toc}{section}{References}
\setboolean{inbibliography}{true}
\bibliographystyle{LHCb}
\bibliography{standard,LHCb-PAPER,LHCb-CONF,LHCb-DP,LHCb-TDR,pHe}

\ifx\mcitethebibliography\mciteundefinedmacro
\PackageError{LHCb.bst}{mciteplus.sty has not been loaded}
{This bibstyle requires the use of the mciteplus package.}\fi
\providecommand{\href}[2]{#2}
\begin{mcitethebibliography}{10}
\mciteSetBstSublistMode{n}
\mciteSetBstMaxWidthForm{subitem}{\alph{mcitesubitemcount})}
\mciteSetBstSublistLabelBeginEnd{\mcitemaxwidthsubitemform\space}
{\relax}{\relax}

\bibitem{Gaisser:1974ks}
T.~K. Gaisser and E.~H. Levy,
  \ifthenelse{\boolean{articletitles}}{\emph{{Astrophysical implications of
  cosmic-ray antiprotons}},
  }{}\href{http://dx.doi.org/10.1103/PhysRevD.10.1731}{Phys.\ Rev.\
  \textbf{D10} (1974) 1731}\relax
\mciteBstWouldAddEndPuncttrue
\mciteSetBstMidEndSepPunct{\mcitedefaultmidpunct}
{\mcitedefaultendpunct}{\mcitedefaultseppunct}\relax
\EndOfBibitem
\bibitem{Steigman:1976ev}
G.~Steigman, \ifthenelse{\boolean{articletitles}}{\emph{{Observational tests of
  antimatter cosmologies}},
  }{}\href{http://dx.doi.org/10.1146/annurev.aa.14.090176.002011}{Ann.\ Rev.\
  Astron.\ Astrophys.\  \textbf{14} (1976) 339}\relax
\mciteBstWouldAddEndPuncttrue
\mciteSetBstMidEndSepPunct{\mcitedefaultmidpunct}
{\mcitedefaultendpunct}{\mcitedefaultseppunct}\relax
\EndOfBibitem
\bibitem{Silk:1984zy}
J.~Silk and M.~Srednicki,
  \ifthenelse{\boolean{articletitles}}{\emph{{Cosmic-ray antiprotons as a probe
  of a photino-dominated Universe}},
  }{}\href{http://dx.doi.org/10.1103/PhysRevLett.53.624}{Phys.\ Rev.\ Lett.\
  \textbf{53} (1984) 624}\relax
\mciteBstWouldAddEndPuncttrue
\mciteSetBstMidEndSepPunct{\mcitedefaultmidpunct}
{\mcitedefaultendpunct}{\mcitedefaultseppunct}\relax
\EndOfBibitem
\bibitem{Stecker:1985jc}
F.~W. Stecker, S.~Rudaz, and T.~F. Walsh,
  \ifthenelse{\boolean{articletitles}}{\emph{{Galactic antiprotons from
  photinos}}, }{}\href{http://dx.doi.org/10.1103/PhysRevLett.55.2622}{Phys.\
  Rev.\ Lett.\  \textbf{55} (1985) 2622}\relax
\mciteBstWouldAddEndPuncttrue
\mciteSetBstMidEndSepPunct{\mcitedefaultmidpunct}
{\mcitedefaultendpunct}{\mcitedefaultseppunct}\relax
\EndOfBibitem
\bibitem{Hagelin:1985pv}
J.~S. Hagelin and G.~L. Kane,
  \ifthenelse{\boolean{articletitles}}{\emph{{Cosmic ray antimatter from
  supersymmetric dark matter}},
  }{}\href{http://dx.doi.org/10.1016/0550-3213(86)90123-9}{Nucl.\ Phys.\
  \textbf{B263} (1986) 399}\relax
\mciteBstWouldAddEndPuncttrue
\mciteSetBstMidEndSepPunct{\mcitedefaultmidpunct}
{\mcitedefaultendpunct}{\mcitedefaultseppunct}\relax
\EndOfBibitem
\bibitem{pamela}
PAMELA collaboration, O.~Adriani {\em et~al.},
  \ifthenelse{\boolean{articletitles}}{\emph{{Measurement of the flux of
  primary cosmic ray antiprotons with energies of 60 MeV to 350 GeV in the
  PAMELA experiment}},
  }{}\href{http://dx.doi.org/10.1134/S002136401222002X}{JETP Letters
  \textbf{96} (2013) 621}\relax
\mciteBstWouldAddEndPuncttrue
\mciteSetBstMidEndSepPunct{\mcitedefaultmidpunct}
{\mcitedefaultendpunct}{\mcitedefaultseppunct}\relax
\EndOfBibitem
\bibitem{ams}
AMS collaboration, M.~Aguilar {\em et~al.},
  \ifthenelse{\boolean{articletitles}}{\emph{{Antiproton flux,
  antiproton-to-proton flux ratio, and properties of elementary particle fluxes
  in primary cosmic rays measured with the Alpha Magnetic Spectrometer on the
  International Space Station}},
  }{}\href{http://dx.doi.org/10.1103/PhysRevLett.117.091103}{Phys.\ Rev.\
  Lett.\  \textbf{117} (2016) 091103}\relax
\mciteBstWouldAddEndPuncttrue
\mciteSetBstMidEndSepPunct{\mcitedefaultmidpunct}
{\mcitedefaultendpunct}{\mcitedefaultseppunct}\relax
\EndOfBibitem
\bibitem{diMauro:2014zea}
M.~di~Mauro, F.~Donato, A.~Goudelis, and P.~D. Serpico,
  \ifthenelse{\boolean{articletitles}}{\emph{{New evaluation of the antiproton
  production cross section for cosmic ray studies}},
  }{}\href{http://dx.doi.org/10.1103/PhysRevD.90.085017}{Phys.\ Rev.\
  \textbf{D90} (2014) 085017},
  \href{http://arxiv.org/abs/1408.0288}{{\normalfont\ttfamily
  arXiv:1408.0288}}\relax
\mciteBstWouldAddEndPuncttrue
\mciteSetBstMidEndSepPunct{\mcitedefaultmidpunct}
{\mcitedefaultendpunct}{\mcitedefaultseppunct}\relax
\EndOfBibitem
\bibitem{Giesen:2015ufa}
G.~Giesen {\em et~al.}, \ifthenelse{\boolean{articletitles}}{\emph{{AMS-02
  antiprotons, at last! Secondary astrophysical component and immediate
  implications for dark matter}},
  }{}\href{http://dx.doi.org/10.1088/1475-7516/2015/09/023}{JCAP \textbf{09}
  (2015) 023}, \href{http://arxiv.org/abs/1504.04276}{{\normalfont\ttfamily
  arXiv:1504.04276}}\relax
\mciteBstWouldAddEndPuncttrue
\mciteSetBstMidEndSepPunct{\mcitedefaultmidpunct}
{\mcitedefaultendpunct}{\mcitedefaultseppunct}\relax
\EndOfBibitem
\bibitem{Kappl:2015bqa}
R.~Kappl, A.~Reinert, and M.~W. Winkler,
  \ifthenelse{\boolean{articletitles}}{\emph{{AMS-02 antiprotons reloaded}},
  }{}\href{http://dx.doi.org/10.1088/1475-7516/2015/10/034}{JCAP \textbf{10}
  (2015) 034}, \href{http://arxiv.org/abs/1506.04145}{{\normalfont\ttfamily
  arXiv:1506.04145}}\relax
\mciteBstWouldAddEndPuncttrue
\mciteSetBstMidEndSepPunct{\mcitedefaultmidpunct}
{\mcitedefaultendpunct}{\mcitedefaultseppunct}\relax
\EndOfBibitem
\bibitem{Reinert:2017aga}
A.~Reinert and M.~W. Winkler, \ifthenelse{\boolean{articletitles}}{\emph{{A
  precision search for WIMPs with charged cosmic rays}},
  }{}\href{http://dx.doi.org/10.1088/1475-7516/2018/01/055}{JCAP \textbf{1801}
  (2018) 055}, \href{http://arxiv.org/abs/1712.00002}{{\normalfont\ttfamily
  arXiv:1712.00002}}\relax
\mciteBstWouldAddEndPuncttrue
\mciteSetBstMidEndSepPunct{\mcitedefaultmidpunct}
{\mcitedefaultendpunct}{\mcitedefaultseppunct}\relax
\EndOfBibitem
\bibitem{Donato:2017ywo}
F.~Donato, M.~Korsmeier, and M.~Di~Mauro,
  \ifthenelse{\boolean{articletitles}}{\emph{{Prescriptions on antiproton cross
  section data for precise theoretical antiproton flux predictions}},
  }{}\href{http://dx.doi.org/10.1103/PhysRevD.96.043007}{Phys.\ Rev.\
  \textbf{D96} (2017) 043007},
  \href{http://arxiv.org/abs/1704.03663}{{\normalfont\ttfamily
  arXiv:1704.03663}}\relax
\mciteBstWouldAddEndPuncttrue
\mciteSetBstMidEndSepPunct{\mcitedefaultmidpunct}
{\mcitedefaultendpunct}{\mcitedefaultseppunct}\relax
\EndOfBibitem
\bibitem{Alves:2008zz}
LHCb collaboration, A.~A. Alves~Jr.\ {\em et~al.},
  \ifthenelse{\boolean{articletitles}}{\emph{{The \lhcb detector at the LHC}},
  }{}\href{http://dx.doi.org/10.1088/1748-0221/3/08/S08005}{JINST \textbf{3}
  (2008) S08005}\relax
\mciteBstWouldAddEndPuncttrue
\mciteSetBstMidEndSepPunct{\mcitedefaultmidpunct}
{\mcitedefaultendpunct}{\mcitedefaultseppunct}\relax
\EndOfBibitem
\bibitem{LHCb-DP-2014-002}
LHCb collaboration, R.~Aaij {\em et~al.},
  \ifthenelse{\boolean{articletitles}}{\emph{{LHCb detector performance}},
  }{}\href{http://dx.doi.org/10.1142/S0217751X15300227}{Int.\ J.\ Mod.\ Phys.\
  \textbf{A30} (2015) 1530022},
  \href{http://arxiv.org/abs/1412.6352}{{\normalfont\ttfamily
  arXiv:1412.6352}}\relax
\mciteBstWouldAddEndPuncttrue
\mciteSetBstMidEndSepPunct{\mcitedefaultmidpunct}
{\mcitedefaultendpunct}{\mcitedefaultseppunct}\relax
\EndOfBibitem
\bibitem{LHCb-DP-2012-003}
M.~Adinolfi {\em et~al.},
  \ifthenelse{\boolean{articletitles}}{\emph{{Performance of the \lhcb RICH
  detector at the LHC}},
  }{}\href{http://dx.doi.org/10.1140/epjc/s10052-013-2431-9}{Eur.\ Phys.\ J.\
  \textbf{C73} (2013) 2431},
  \href{http://arxiv.org/abs/1211.6759}{{\normalfont\ttfamily
  arXiv:1211.6759}}\relax
\mciteBstWouldAddEndPuncttrue
\mciteSetBstMidEndSepPunct{\mcitedefaultmidpunct}
{\mcitedefaultendpunct}{\mcitedefaultseppunct}\relax
\EndOfBibitem
\bibitem{smog}
C.~Barschel, {\em {Precision luminosity measurement at LHCb with beam-gas
  imaging}}, PhD thesis, RWTH Aachen U., 2014,
  \href{https://cds.cern.ch/record/1693671}{CERN-THESIS-2013-301}\relax
\mciteBstWouldAddEndPuncttrue
\mciteSetBstMidEndSepPunct{\mcitedefaultmidpunct}
{\mcitedefaultendpunct}{\mcitedefaultseppunct}\relax
\EndOfBibitem
\bibitem{LHCb-PAPER-2014-047}
LHCb collaboration, R.~Aaij {\em et~al.},
  \ifthenelse{\boolean{articletitles}}{\emph{{Precision luminosity measurements
  at LHCb}}, }{}\href{http://dx.doi.org/10.1088/1748-0221/9/12/P12005}{JINST
  \textbf{9} (2014) P12005},
  \href{http://arxiv.org/abs/1410.0149}{{\normalfont\ttfamily
  arXiv:1410.0149}}\relax
\mciteBstWouldAddEndPuncttrue
\mciteSetBstMidEndSepPunct{\mcitedefaultmidpunct}
{\mcitedefaultendpunct}{\mcitedefaultseppunct}\relax
\EndOfBibitem
\bibitem{Evans:2008zzb}
L.~Evans and P.~Bryant, \ifthenelse{\boolean{articletitles}}{\emph{{LHC
  Machine}}, }{}\href{http://dx.doi.org/10.1088/1748-0221/3/08/S08001}{JINST
  \textbf{3} (2008) S08001}\relax
\mciteBstWouldAddEndPuncttrue
\mciteSetBstMidEndSepPunct{\mcitedefaultmidpunct}
{\mcitedefaultendpunct}{\mcitedefaultseppunct}\relax
\EndOfBibitem
\bibitem{Pierog:2013ria}
T.~Pierog {\em et~al.}, \ifthenelse{\boolean{articletitles}}{\emph{{EPOS LHC:
  test of collective hadronization with data measured at the CERN Large Hadron
  Collider}}, }{}\href{http://dx.doi.org/10.1103/PhysRevC.92.034906}{Phys.\
  Rev.\  \textbf{C92} (2015) 034906},
  \href{http://arxiv.org/abs/1306.0121}{{\normalfont\ttfamily
  arXiv:1306.0121}}\relax
\mciteBstWouldAddEndPuncttrue
\mciteSetBstMidEndSepPunct{\mcitedefaultmidpunct}
{\mcitedefaultendpunct}{\mcitedefaultseppunct}\relax
\EndOfBibitem
\bibitem{Gramolin:2014pva}
A.~V. Gramolin {\em et~al.}, \ifthenelse{\boolean{articletitles}}{\emph{{A new
  event generator for the elastic scattering of charged leptons on protons}},
  }{}\href{http://dx.doi.org/10.1088/0954-3899/41/11/115001}{J.\ Phys.\
  \textbf{G41} (2014) 115001},
  \href{http://arxiv.org/abs/1401.2959}{{\normalfont\ttfamily
  arXiv:1401.2959}}\relax
\mciteBstWouldAddEndPuncttrue
\mciteSetBstMidEndSepPunct{\mcitedefaultmidpunct}
{\mcitedefaultendpunct}{\mcitedefaultseppunct}\relax
\EndOfBibitem
\bibitem{Allison:2006ve}
Geant4 collaboration, J.~Allison {\em et~al.},
  \ifthenelse{\boolean{articletitles}}{\emph{{Geant4 developments and
  applications}}, }{}\href{http://dx.doi.org/10.1109/TNS.2006.869826}{IEEE
  Trans.\ Nucl.\ Sci.\  \textbf{53} (2006) 270}\relax
\mciteBstWouldAddEndPuncttrue
\mciteSetBstMidEndSepPunct{\mcitedefaultmidpunct}
{\mcitedefaultendpunct}{\mcitedefaultseppunct}\relax
\EndOfBibitem
\bibitem{Agostinelli:2002hh}
Geant4 collaboration, S.~Agostinelli {\em et~al.},
  \ifthenelse{\boolean{articletitles}}{\emph{{Geant4: A simulation toolkit}},
  }{}\href{http://dx.doi.org/10.1016/S0168-9002(03)01368-8}{Nucl.\ Instrum.\
  Meth.\  \textbf{A506} (2003) 250}\relax
\mciteBstWouldAddEndPuncttrue
\mciteSetBstMidEndSepPunct{\mcitedefaultmidpunct}
{\mcitedefaultendpunct}{\mcitedefaultseppunct}\relax
\EndOfBibitem
\bibitem{LHCb-PROC-2011-006}
M.~Clemencic {\em et~al.}, \ifthenelse{\boolean{articletitles}}{\emph{{The
  \lhcb simulation application, Gauss: Design, evolution and experience}},
  }{}\href{http://dx.doi.org/10.1088/1742-6596/331/3/032023}{J.\ Phys.\ Conf.\
  Ser.\  \textbf{331} (2011) 032023}\relax
\mciteBstWouldAddEndPuncttrue
\mciteSetBstMidEndSepPunct{\mcitedefaultmidpunct}
{\mcitedefaultendpunct}{\mcitedefaultseppunct}\relax
\EndOfBibitem
\bibitem{LHCB-DP-2013-002}
LHCb collaboration, R.~Aaij {\em et~al.},
  \ifthenelse{\boolean{articletitles}}{\emph{{Measurement of the track
  reconstruction efficiency at LHCb}},
  }{}\href{http://dx.doi.org/10.1088/1748-0221/10/02/P02007}{JINST \textbf{10}
  (2015) P02007}, \href{http://arxiv.org/abs/1408.1251}{{\normalfont\ttfamily
  arXiv:1408.1251}}\relax
\mciteBstWouldAddEndPuncttrue
\mciteSetBstMidEndSepPunct{\mcitedefaultmidpunct}
{\mcitedefaultendpunct}{\mcitedefaultseppunct}\relax
\EndOfBibitem
\bibitem{LHCb-PAPER-2011-037}
LHCb collaboration, R.~Aaij {\em et~al.},
  \ifthenelse{\boolean{articletitles}}{\emph{{Measurement of prompt hadron
  production ratios in $\proton\proton$ collisions at $\sqrt{s}=0.9$ and
  $7$\tev}}, }{}\href{http://dx.doi.org/10.1140/epjc/s10052-012-2168-x}{Eur.\
  Phys.\ J.\  \textbf{C72} (2012) 2168},
  \href{http://arxiv.org/abs/1206.5160}{{\normalfont\ttfamily
  arXiv:1206.5160}}\relax
\mciteBstWouldAddEndPuncttrue
\mciteSetBstMidEndSepPunct{\mcitedefaultmidpunct}
{\mcitedefaultendpunct}{\mcitedefaultseppunct}\relax
\EndOfBibitem
\bibitem{AdaBoost}
Y.~Freund and R.~E. Schapire, \ifthenelse{\boolean{articletitles}}{\emph{A
  decision-theoretic generalization of on-line learning and an application to
  boosting}, }{}\href{http://dx.doi.org/10.1006/jcss.1997.1504}{J.\ Comput.\
  Syst.\ Sci.\  \textbf{55} (1997) 119}\relax
\mciteBstWouldAddEndPuncttrue
\mciteSetBstMidEndSepPunct{\mcitedefaultmidpunct}
{\mcitedefaultendpunct}{\mcitedefaultseppunct}\relax
\EndOfBibitem
\bibitem{Pierog:2009zt}
T.~Pierog and K.~Werner, \ifthenelse{\boolean{articletitles}}{\emph{{EPOS model
  and ultra high energy cosmic rays}},
  }{}\href{http://dx.doi.org/10.1016/j.nuclphysbps.2009.09.017}{Nucl.\ Phys.\
  Proc.\ Suppl.\  \textbf{196} (2009) 102},
  \href{http://arxiv.org/abs/0905.1198}{{\normalfont\ttfamily
  arXiv:0905.1198}}\relax
\mciteBstWouldAddEndPuncttrue
\mciteSetBstMidEndSepPunct{\mcitedefaultmidpunct}
{\mcitedefaultendpunct}{\mcitedefaultseppunct}\relax
\EndOfBibitem
\bibitem{Gyulassy:1994ew}
M.~Gyulassy and X.-N. Wang, \ifthenelse{\boolean{articletitles}}{\emph{{HIJING
  1.0: A Monte Carlo program for parton and particle production in high-energy
  hadronic and nuclear collisions}},
  }{}\href{http://dx.doi.org/10.1016/0010-4655(94)90057-4}{Comput.\ Phys.\
  Commun.\  \textbf{83} (1994) 307},
  \href{http://arxiv.org/abs/nucl-th/9502021}{{\normalfont\ttfamily
  arXiv:nucl-th/9502021}}\relax
\mciteBstWouldAddEndPuncttrue
\mciteSetBstMidEndSepPunct{\mcitedefaultmidpunct}
{\mcitedefaultendpunct}{\mcitedefaultseppunct}\relax
\EndOfBibitem
\bibitem{Ostapchenko:2010vb}
S.~Ostapchenko, \ifthenelse{\boolean{articletitles}}{\emph{{Monte Carlo
  treatment of hadronic interactions in enhanced Pomeron scheme: QGSJET-II
  model}}, }{}\href{http://dx.doi.org/10.1103/PhysRevD.83.014018}{Phys.\ Rev.\
  \textbf{D83} (2011) 014018},
  \href{http://arxiv.org/abs/1010.1869}{{\normalfont\ttfamily
  arXiv:1010.1869}}\relax
\mciteBstWouldAddEndPuncttrue
\mciteSetBstMidEndSepPunct{\mcitedefaultmidpunct}
{\mcitedefaultendpunct}{\mcitedefaultseppunct}\relax
\EndOfBibitem
\bibitem{Kachelriess:2015wpa}
M.~Kachelriess, I.~V. Moskalenko, and S.~S. Ostapchenko,
  \ifthenelse{\boolean{articletitles}}{\emph{{New calculation of antiproton
  production by cosmic ray protons and nuclei}},
  }{}\href{http://dx.doi.org/10.1088/0004-637X/803/2/54}{Astrophys.\ J.\
  \textbf{803} (2015) 54},
  \href{http://arxiv.org/abs/1502.04158}{{\normalfont\ttfamily
  arXiv:1502.04158}}\relax
\mciteBstWouldAddEndPuncttrue
\mciteSetBstMidEndSepPunct{\mcitedefaultmidpunct}
{\mcitedefaultendpunct}{\mcitedefaultseppunct}\relax
\EndOfBibitem
\bibitem{Sjostrand:2006za}
T.~Sj\"{o}strand, S.~Mrenna, and P.~Skands,
  \ifthenelse{\boolean{articletitles}}{\emph{{PYTHIA 6.4 physics and manual}},
  }{}\href{http://dx.doi.org/10.1088/1126-6708/2006/05/026}{JHEP \textbf{05}
  (2006) 026}, \href{http://arxiv.org/abs/hep-ph/0603175}{{\normalfont\ttfamily
  arXiv:hep-ph/0603175}}\relax
\mciteBstWouldAddEndPuncttrue
\mciteSetBstMidEndSepPunct{\mcitedefaultmidpunct}
{\mcitedefaultendpunct}{\mcitedefaultseppunct}\relax
\EndOfBibitem
\bibitem{Korsmeier:2018gcy}
M.~Korsmeier, F.~Donato, and M.~Di~Mauro,
  \ifthenelse{\boolean{articletitles}}{\emph{{Production cross sections of
  cosmic antiprotons in the light of new data from the NA61 and LHCb
  experiments}}, }{}\href{http://dx.doi.org/10.1103/PhysRevD.97.103019}{Phys.\
  Rev.\ D \textbf{97} (2018) 103019},
  \href{http://arxiv.org/abs/1802.03030}{{\normalfont\ttfamily
  arXiv:1802.03030}}\relax
\mciteBstWouldAddEndPuncttrue
\mciteSetBstMidEndSepPunct{\mcitedefaultmidpunct}
{\mcitedefaultendpunct}{\mcitedefaultseppunct}\relax
\EndOfBibitem
\end{mcitethebibliography}

\newpage


\clearpage

\centerline{\large\bf LHCb collaboration}
\begin{flushleft}
\small
R.~Aaij$^{28}$,
C.~Abell{\'a}n~Beteta$^{45}$,
B.~Adeva$^{42}$,
M.~Adinolfi$^{49}$,
C.A.~Aidala$^{74}$,
Z.~Ajaltouni$^{5}$,
S.~Akar$^{60}$,
P.~Albicocco$^{19}$,
J.~Albrecht$^{10}$,
F.~Alessio$^{43}$
% ...

{\footnotesize \it
% ...
$ ^{5}$Clermont Universit{\'e}, Universit{\'e} Blaise Pascal, CNRS/IN2P3, LPC, Clermont-Ferrand, France\\
% ...
$ ^{10}$Fakult{\"a}t Physik, Technische Universit{\"a}t Dortmund, Dortmund, Germany\\
% ...
$ ^{19}$INFN Laboratori Nazionali di Frascati, Frascati, Italy\\
% ...
$ ^{28}$Nikhef National Institute for Subatomic Physics, Amsterdam, Netherlands\\
% ...
$ ^{42}$Instituto Galego de F{\'\i}sica de Altas Enerx{\'\i}as (IGFAE), Universidade de Santiago de Compostela, Santiago de Compostela, Spain\\
$ ^{43}$European Organization for Nuclear Research (CERN), Geneva, Switzerland\\
% ...
$ ^{45}$Physik-Institut, Universit{\"a}t Z{\"u}rich, Z{\"u}rich, Switzerland\\
% ...
$ ^{49}$H.H. Wills Physics Laboratory, University of Bristol, Bristol, United Kingdom\\
% ...
$ ^{60}$University of Cincinnati, Cincinnati, OH, United States\\
% ...
$ ^{62}$Syracuse University, Syracuse, NY, United States\\
% ...
$ ^{74}$University of Michigan, Ann Arbor, United States, associated to $^{62}$\\
% ...
}
\end{flushleft}

\end{document}